\documentclass[apj]{emulateapj}

\def\baselinestretch{0.96}

\begin{document}

\title{Protoplanetary disks including radiative feedback from accreting planets}

\author{Mat\'ias Montesinos$^{1,2}$, Jorge Cuadra$^{2}$, Sebastian Perez$^{1}$, Cl\'ement Baruteau$^{3}$, Simon Casassus$^{1}$}

\affil{$^{1}$Departamento de Astronom\'ia, Universidad de Chile,
Casilla 36-D, Santiago, Chile; montesinos@das.uchile.cl}

\affil{$^{2}$Instituto de Astrof\'isica, Pontificia Universidad Cat\'olica de Chile, Santiago, Chile}

\affil{$^{3}$Institut de Recherche en Astrophysique et Plan\'etologie, CNRS / Universit\'e de Toulouse / UPS-OMP, 14 avenue Edouard Belin, 31400 Toulouse, France.}

%
%
%
%
%



\begin{abstract}
  While recent observational progress is converging on the detection
  of compact regions of thermal emission due to embedded protoplanets,
  further theoretical predictions are needed to understand the
  response of a protoplanetary disk to the planet formation radiative
  feedback. This is particularly important to make predictions for the
  observability of  circumplanetary regions. In this work we
  use 2D hydrodynamical simulations to examine the evolution of a
  viscous protoplanetary disk in which a luminous Jupiter-mass planet
  is embedded. We use an energy equation which includes the radiative
  heating of the planet as an additional mechanism for planet
  formation feedback. Several models are computed for planet
  luminosities ranging from $10^{-5}$ to $10^{-3}$ Solar
  luminosities. We find that the planet radiative feedback enhances
  the disk's accretion rate at the planet's orbital radius, producing
  a hotter and more luminous environement around the planet,
  independently of the prescription used to model the disk's turbulent
  viscosity. We also estimate the thermal signature of the planet
  feedback for our range of planet luminosities, finding that the
  emitted spectrum of a purely active disk, without passive heating,
  is appreciably modified in the infrared. We simulate the
  protoplanetary disk around HD 100546 where a planet companion is
  located at about 68 au from the star. Assuming the planet mass is 5
  Jupiter masses and its luminosity is $\sim 2.5 \times 10^{-4} \,
  L_\odot$, we find that the radiative feedback of the planet
  increases the luminosity of its $\sim 5$ au circumplanetary disk
  from $10^{-5} \, \rm L_\odot$ (without feedback) to
  $10^{-3} \, \rm L_\odot$, corresponding to an emission of $\sim 1 \,
  \rm mJy$ in $L^\prime$ band after radiative transfer calculations, a
  value that is in good agreement with HD 100546b observations.
  \end{abstract}

\keywords{planet-disk interactions - protoplanetary disks - accretion,
  accretion disks - planetary systems - hydrodynamics - methods:
  numerical}


\section{Introduction}

Recent observational progress allows the detailed study of planet
formation feedback.  The ALMA facility has opened the resolved study
of accretion kinematics in protoplanetary gaps. For instance, Casassus et al. (\cite{Casassus-et-al-2012}, \cite{Casassus-et-al-2013}) find that the dust gap in the disk around the
star HD 142527 shows a disrupted outer disk suggestive of on-going
dynamical clearing, and contains residual gas whose kinematics are
consistent with accretion across the dust gap.  Dramatic advances in
high-contrast imaging techniques have allowed the likely detection of
embedded protoplanets. \cite{Quanz-et-al-2013b} found a compact but
resolved $3.8 \mu\rm m$ ($L^\prime$) source at $\sim 68 \, \rm au$
from HD 100546 (independently confirmed by \cite{Currie-et-al-2014}, which could be interpreted as on-going accretion onto a compact body. In
another example of resolved data, \cite{Quanz-et-al-2013a} present
polarized light images of HD 169142 resolving features in its
protoplanetary disk that could be interpreted as a gap induced by
forming protoplanets. Indeed, \cite{Reggiani-et-al-2014} and \cite{Biller-et-al-2014} find an $L^\prime$ point source, within this gap, at a
separation of $\sim 22~$au. There are indications that this $L^\prime$
compact signal is not photospheric, and that it is somehow connected
to the protoplanetary accretion luminosity. Similar findings have been
reported for the compact H$\alpha$ signal found at 12~au from HD
142527 by \cite{Close-et-al-2014}, which coincides with a relatively
bright $L^\prime$ signal. That source would reach the stellar mass
regime for photospheric emission \citep{Biller-et-al-2014}, but has been
recently resolved to be extended and polarized in $Y$ band \citep{Rodigas-et-al-2014}, suggesting that this companion to HD 142527 is probably a
substellar object with a remarkably strong thermal luminosity in
$L^\prime$, that is somehow connected to accretion.

Hydrodynamical simulations can be used to inform the interpretation of
the data on planet forming systems, by calculating the radiative
emissions of a disk with an embedded planet that has created a gap.
This was the approach used by \cite{Wolf-D'Angelo-2005}, who modelled in
two dimensions the gravitational response of a disk to the presence of
an embedded planet, under the assumption of a constant aspect ratio
for the disk.  As a post-processing step, once the simulation reached
steady state, \cite{Wolf-D'Angelo-2005} assumed a luminosity for the planet and calculated the radiative response of the disk.  They concluded that
 a hot circumplanetary region could
eventually be detected by ALMA.

In this paper we also use 2D hydrodynamical simulations to follow the
dynamics of a disk with an embedded planet.  However, instead of
assuming a temperature profile for the disk, we use a non-stationary
energy equation that includes the radiative feedback of planet
formation and a temperature-dependent black body cooling for the disk.

The structure of the paper is as follows: In \S\ref{model} we present
the model including the main assumptions, the physical conditions of
the disk, as well as the numerical setup and a description of the
code. Our results and main conclusions of the evolution of the density
profile, temperature, and the spectral signature of the disk are
presented in \S\ref{results}, with a short discussion in
\S\ref{discussion}.  We summarize our findings in \S\ref{summary}.

\section{The model}\label{model}

We are interested in the evolution of a gaseous protoplanetary disk in
which a luminous Jupiter-mass planet is embedded. In our simulations
we use an energy equation that includes radiative cooling, and both
viscous heating and heating due to the planet's luminosity.  We follow
the evolution of the protoplanetary disk for about $10^4$ years,
assuming the planet is already formed at the beginning of the
simulations. This is a short period of time compared with both the lifetime
of protoplanetary disks ($\sim \rm Myr$ e.g., \cite{Williams-Cieza-2011}), and the time-scales over which planet
luminosities should vary (\cite{Marley-et-al-2007}, \cite{Mordasini-2013}), thus justifying the use of a constant planet luminosity in
the simulations.

Our simulations use the public two-dimensional hydrodynamics code
FARGO-AD\footnote{http://fargo.in2p3.fr/spip.php?rubrique9} \citep{Baruteau-Masset-2008} which is dedicated to planet--disk interactions.  It
is a staggered mesh code that solves the Navier-Stokes, continuity and
energy equations on a polar grid. It is based on an Eulerian formalism
using a finite difference method of second order, according to the \cite{VanLeer-1977} upwind algorithm. Details of the code can be found in
\cite{Masset-2000} and \cite{Baruteau-Masset-2008}. In FARGO-AD's public
version, the energy equation includes viscous heating and a simple
temperature relaxation to reach thermodynamical equilibrium over some 
(user-defined) characteristic timescale.

The present work features two main changes to the energy equation. One is the inclusion of a radiative cooling function, based on the assumption
that the disk radiates locally as a blackbody. The second, and most
important feature, is the implementation of a heating source term
associated to the planet. We assume that the protoplanet has an
intrinsic constant luminosity, therefore, it injects energy into the
disk at a constant rate.  We only take into account the
gravitational potentials of the star and of the planet, the disk's
self-gravity is neglected.

\subsection{Code units and Initial Setup} \label{units}

We set the mass of the central star ($M_{\star}$) and the planet's
orbital radius ($r_{\rm p}$) as the code's units of mass and
length, respectively.  The code's unit of time ($t_0$) is the planet's
orbital period divided by $2\pi$, that is $t_0=(G M_{\star} / r_{\rm
  p}^3)^{-1/2}$. The gravitational constant $G=1$ in code units. The
code's unit of temperature is $G M_{\star} \mu m_p / (k_B r_{\rm p})$,
with $\mu$ the mean molecular weight of the gas ($\mu = 2.35$
in all our simulations), $m_p$ the proton mass and $k_B$ the Boltzmann
constant. Unless otherwise noted (see \S~\ref{sec:HD100546}), 
we adopt a solar-mass star ($M_{\star}=M_\odot$) and a planet at 
$r_{\rm p} = 10\, \rm au$.

We use cylindrical coordinates $(r,\phi)$. The computational domain
extends from $r = 1$ to $50 \, \rm au$ over $n_r = 400$ equally spaced
radial rings. It covers the full $2\pi$ extent in azimuth over $n_\phi
= 800$ equally spaced azimuthal sectors. Tests with higher grid
resolutions,$n_r \times n_\phi = 512 \times 1536$, were performed to check the convergence of  our results. We use an open inner and outer boundary conditions, meaning that the material is allowed to outflow at the disk edges.

The initial density profile scales with $r^{-1}$: 
\begin{equation}
\Sigma(r) = \Sigma_0 \frac{r_p}{r},
\label{sigmaScale}
\end{equation}
where $\Sigma_0 = 2.56 \times 10^{-5} \, \rm M_\odot/au^{2}$. This
initial disk mass is thus $10^{-3} \rm M_\odot$. The disk's aspect
ratio $h = c_{\rm s} / v_{\rm K}$, with $c_{\rm s}$ the isothermal
sound speed and $v_{\rm K}$ the Keplerian velocity, initially equals
0.05, uniformly. The disk's initial temperature therefore decreases in
$r^{-1}$ and $\approx 63$ K at 10 au for our fiducial primary mass
($M_{\star} = M_\odot$).

We fix the planet-to-primary mass ratio ($q$) to $q=10^{-3}$, so the
planet has a Jovian mass for a Solar-mass star. The planet is held on
a fixed circular orbit (it does not migrate through the disk). To
avoid a violent response of the disk to the planet's gravitational
potential initially, the planet mass is increased gradually over the
first five orbits according to
\begin{equation}
M(t) = M_{\rm p} \sin^2(\pi t / 10 T_{\rm p}),
\label{Tapper}
\end{equation}
where $T_{\rm p}$ is the planet's orbital period. We adopt three values 
for the planet luminosity: $10^{-5}$, $10^{-4}$, and $10^{-3}$
$\rm L_\odot$ (see \S~\ref{planet-feedback}). For the viscosity
prescription, in our fiducial model we use an alpha disk model
\cite{Shakura-Sunyaev-1973} setting $\alpha = 4 \times 10^{-3}$ (but see
\S~\ref{viscosities}).

\subsection{The energy equation}

The energy equation satisfied by the thermal energy density $e$ reads
(e.g., cite{D'Angelo-et-al-2003})
\begin{equation}
\frac{\partial e}{\partial t} + \overrightarrow{\nabla} \cdot (e \overrightarrow{v}) = -P \overrightarrow{\nabla} \cdot \overrightarrow{v} + 
Q^+ - Q^-,
\label{energy2}
\end{equation}
where $\overrightarrow{v}$ is the gas velocity, $P$ the pressure,
$Q^+$ the heating rate per unit area, and $Q^-$ the radiative cooling
rate per unit area. To close the system of equations, an ideal
equation of state is used,
\begin{equation}
P = \Sigma  T  \overline{R} ,
\label{state1}
\end{equation} 
with $T$ the gas temperature and $\overline{R} = k_B / \mu m_p $. The
thermal energy density is related to the temperature through
\begin{equation}
e = \Sigma T \left(  \frac{\overline{R}}{ \gamma - 1}  \right),
\label{state2}
\end{equation}
where $\gamma$ denotes the adiabatic index, which we fix to $\gamma =
1.4$ (a typical value for a diatomic gas). Eq. (\ref{energy2}) can be
recast as
\begin{eqnarray}
\frac{\partial e}{\partial t} + \overrightarrow{\nabla} \cdot (e \overrightarrow{v}) = -(\gamma - 1) e \overrightarrow{\nabla} \cdot \overrightarrow{v} ~+~{} \nonumber\\
{}   Q_{\rm v}^+ +Q_{\rm p}^+ - Q^-,
\label{energy3}
\end{eqnarray}
where $Q_{\rm v}^+$ is the viscous heating rate, $Q_{\rm p}^+$ the
flux of radiative energy received from the planet (feedback), and
$Q^-$ corresponds to the radiative cooling rate of the disk. These
source terms are detailed in the next section.

In a more realistic situation, a thermal diffusion
flux term  should be added on the right side of Eq. (\ref{energy3}).
But, since the outer temperature of the CPD  matches that of the protoplanetary disk, we expect no strong heat diffusion  to occur.

\subsection{Sources of heating and cooling}

\subsubsection{Viscous dissipation}
\label{sec:viscousdiss}

The viscous heating rate implemented in FARGO-AD, $Q^+_{\rm v}$, has
two contributions. The first one arises from the shear kinematic viscosity
$\nu$,
\begin{equation}
  Q^+_{\rm shear} = \frac{1}{2 \nu \Sigma} \left[ \tau^2_{r,r} + 2 \tau^2_{r,\phi} + \tau^2_{\phi,\phi} \right]
  + \frac{2 \nu \Sigma}{9} (\overrightarrow{\nabla}\cdot\overrightarrow{v})^2,
\label{Qv}
\end{equation}
where the $\tau_{\alpha, \beta}$ are the components of the viscous stress tensor:
\begin{eqnarray}
  \tau_{r,r} &=& 2 \nu \Sigma  \left[ \frac{\partial v_r}{\partial r} - \frac{1}{3}  \overrightarrow{\nabla} \cdot  \overrightarrow{v}  \right],      \nonumber \\
  \tau_{r,\phi}  &=& \nu \Sigma \left[ r \frac{\partial}{\partial r}   \left( \frac{v_\phi}{r} \right)  + \frac{1}{r} \frac{\partial v_r}{\partial \phi}   \right],   \nonumber \\
  \tau_{\phi,\phi}     & = & 2 \nu \Sigma \left[ \frac{1}{r} \frac{\partial v_\phi}{\partial \phi} + \frac{v_r}{r} - \frac{1}{3} \overrightarrow{\nabla} \cdot \overrightarrow{v}  \right].
\label{ViscousTensor}
\end{eqnarray}
Note that for a Keplerian disk
(i.e., $v_r = 0$, $v_\phi = r \Omega_K$ with $\Omega_K$ the Keplerian
angular frequency), $\tau_{r,r}=\tau_{\phi,\phi}=0$ and the shear viscous
heating rate reduces to $Q^+_{\rm v} = \tau^2_{r,\phi}/\nu \Sigma =
\frac{9}{4} \nu \Sigma \Omega^2_K$.

The second contribution to the viscous heating rate is through the use
of a von Neumann-Richtmyer artificial bulk viscosity, as described in
\cite{Stone-Norman-1992}, where the coefficient $C_2$ is taken equal to
1.4 ($C_2$ measures the number of zones over which a shock is spread
over by the artificial viscosity).  

\subsubsection{Planet feedback} \label{planet-feedback}

The key feature in this work is the inclusion of the planet feedback
(i.e., the energy flux received from the protoplanet by the gas disk)
during its evolution. In this initial study, we use a simplified model
for the planet luminosity, which is powered by its mass build up
during its formation and evolution.  Assuming free-fall, the typical
power dissipated during the formation of a Jupiter-like planet is
\begin{equation}
\dot{E}_{\rm J} = \frac{G M_{\rm J}^2}{R_{\rm J} t_{\rm p}},
\label{edot}
\end{equation}
where $M_{\rm J}$ and $R_{\rm J}$ are the mass and radius of Jupiter,
respectively, and $t_{\rm p}$ is a characteristic timescale for its
formation.

A characteristic timescale obtained from ground-based and
\textit{Spitzer}-based infrared (IR) surveys of young stellar
clusters, which trace the evolution of primordial protoplanetary
disks, suggest a formation period of about $\sim 3 \times 10^6$ $\rm
yr$ \citep{Mamajek-2009}. Then, the emitted energy (Eq. \ref{edot})
reads $\dot{E}_{\rm J} \simeq 1 \times 10^{-4}$ $\rm L_\odot$, which
also agrees with hot accretion shock structure formation models (e.g.,
\cite{Mordasini-2013}). To account for the large uncertainties  in the
accretion process of planetesimals, we adopt planet luminosities
$L_{\rm p}$ ranging from $10^{-1} \dot{E}_{\rm J}$ to $10 \dot{E}_{\rm
  J}$ (i.e., $10^{-5}$ - $10^{-3} ~ \rm L_\odot$). 
  
In our model, the planet has already reached its final mass at the
 beginning of the simulation, and no longer grows. In that sense the source
 of the planet luminosity should be thought of as a ``post-formation" slow contraction of the planet, rather than accretion luminosity.  However,  it is still possible that with  opacities larger than those commonly used (e.g., \cite{Bell-Lin-1994}) the accretion shock energy takes longer to be released, and therefore the total planet luminosity might include contributions of both aforementioned  sources. Regarding the luminosities, it should be noted that the high values we are using (i.e., $L_p = 10^{-3} ~ \rm L_\odot$) are expected for large opacities  under the  assumption of a hot accretion formation model \citep{Mordasini-2013}. 
 This is different with cold accretion initial conditions, which would
produce smaller post-formation planet luminosities ($\lesssim 10^{-4} ~ \rm L_\odot$).  As current models of giant planet formation
cannot distinguish between them, we assume hot accretion and study the effect of high planet luminosities.

In all our simulations, the planet's circumplanetary disk (CPD) is
optically thick. The thermal energy released by the planet is
therefore chosen to be distributed isotropically within the planet's
CPD, the size of which is denoted by $R_{\rm CPD}$. Following \cite{Crida-et-al-2009}, we adopt $R_{\rm CPD} = 0.6 R_{\rm Hill}$, where $R_{\rm
  Hill} = r_{\rm p} (q/3)^{1/3}$ is the planet's Hill radius. To avoid
possible thermal shocks at the beginning of the simulation, we
gradually inject the planet energy in the same way as we do for the
planet mass (see Eq. \ref{Tapper}) i.e., following $Q^+_{\rm p}(t) =
Q^+_{\rm p} \sin^2(\pi t /10 T_{\rm p})$, with $T_{\rm p}$ the
planet's orbital period, and
\begin{equation}
Q^+_{\rm p} = \left\{ \begin{array}{ll}
 f(r) L_{\rm p} /(\pi R_{\rm CPD}^2) &\mbox{ if $r=|\overrightarrow{r}-\overrightarrow{r_{\rm p}}|<R_{\rm CPD}$} \\
  0 &\mbox{ otherwise},
       \end{array} \right.
\label{Qp}
\end{equation}
where $f(r)$ is a Gaussian function used to smoothly inject the planet
energy within the CPD. Its expression is $f(r) = A \exp(-5d^2/R_{\rm
  CPD}^2)$, with $A$ is a dimensionless normalization factor equal to
5. It has a FWHM of $\sim 0.75 R_{\rm CPD}$.

It is worth noting that our model has inherent limitations due to its 2D geometry. For instance,  the injected energy from the planet is only transported  through the $r$-$\phi$ plane of the CPD, neglecting the energy escape in the vertical direction. This could lead to an overestimation of the thermal energy deposited on the disk.

\subsubsection{Radiative cooling and disk spectrum}

The cooling term $Q^-$ in Eq.~(\ref{energy3}) corresponds to the
energy flux radiated by the disk in the vertical direction. This
quantity depends on whether the disk is optically thin or
thick. Several processes could be responsible for the energy
evacuation, e.g., for high temperatures ($\sim 10^4 \rm K$) Thomson
scattering processes, free-free, bound-free transitions are
dominant. For lower temperatures e.g., $\sim 1 - 10^3 \, \rm K$, as in
our case, Rosseland and Planck mean opacities of dust and grain
species should be used (e.g., \cite{Bell-Lin-1994}, \cite{Semenov-et-al-2003}).

In our models, we calculate the optical depth ($\tau = \kappa
\Sigma/2$) assuming the Rosseland mean opacity $\kappa$ from \cite{Bell-Lin-1994}.  In some regions of the disk, the gas density could be
diluted (e.g., at the gap, and/or planet location), and become
transparent or less opaque ($\tau \sim 1$). We use the \cite{Hubeny-1990}
prescription to calculate the effective optical depth,

\begin{equation}
\tau_{\rm eff} = \frac{3 \tau}{8} + \frac{\sqrt{3}}{4} + \frac{1}{4 \tau}.
\end{equation}

The effective temperature ($T_{\rm eff}$) and mid-plane temperature ($T$) are 
then related through
\begin{equation}
T^4_{\rm eff} = \frac{T^4}{\tau_{\rm eff}}.
\end{equation}

The cooling rate per unit area due to radiation from the surface of
the disk is calculated by integrating the total emitted radiative flux
$\Phi_\nu$ over all frequencies,
 \begin{equation}
 Q^-(r) = 2 \int_0^\infty \Phi_\nu(T_{\rm eff}(r,\phi)) d\nu,
 \label{Q-}
 \end{equation}
 where $\Phi_\nu(T_{\rm eff}(r,\phi))$ is the local emergent flux of
 the disk surface. The factor 2 is needed because radiation escapes
 from both sides (top and bottom) of the disk. For simplicity, we
 assume that the local emitted flux is given by the blackbody
 approximation,
\begin{equation}
\Phi_\nu(T) = 2 \pi B_\nu(T_{\rm eff}(r,\phi)),
\label{Phi_nu}
\end{equation}
where $B_\nu(T_{\rm eff}(r,\phi))$ is the Planck function, and $T_{\rm
  eff}(r,\phi)$ is the surface disk temperature.

The spectrum of the disk is calculated by integrating the emitted flux
$\Phi_\nu(T_{\rm eff}(r,\phi))$ (Eq. \ref{Phi_nu}) over the surface of
the disk,
\begin{equation}
L_\nu = \int \Phi_\nu(T_{\rm eff}(r,\phi)) r dr d\phi.
\label{spec1}
\end{equation}
Using Eq.~(\ref{spec1}) we calculate the bolometric luminosity of the
disk integrating over all the frequency domain,
\begin{equation}
L_{\rm b} = \int_0^\infty L_\nu d\nu.
\label{Bolometric}
\end{equation}
It corresponds to the electromagnetic energy per unit time radiated
away in all wavelength (without including irradiation from the star).

\section{Results}\label{results} 

In this section we present results of simulations for three different
planet luminosities: $L_{\rm p} = 10^{-5}$, $10^{-4}$, and $10^{-3}$
$\rm L_\odot$. We pay special attention to the gas properties (e.g.,
density, temperature) and to the spectral consequences of planet
feedback for the highest planet luminosity, i.e. $L_{\rm p} = 10^{-3}
\, \rm L_\odot$.

\subsection{Density field}
\label{sec:dens}
\begin{figure} [h]
\plotone{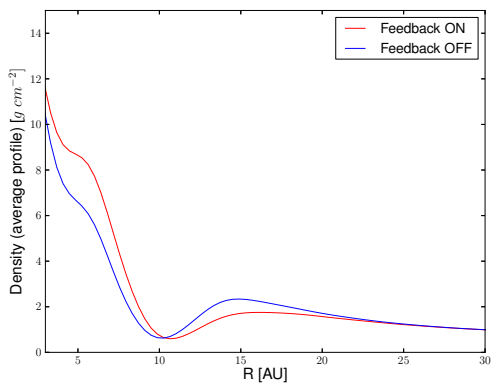}
\caption{Azimuthally-averaged density profiles without and with planet
  feedback (the planet's luminosity is $L_{\rm p}=10^{-3} \, \rm
  L_\odot$).  Both profiles are displayed after 300 orbits of the
  planet. Notice that the inner disk accumulates more material when
  feedback is activated.}
\label{SigmaProfile2}
\end{figure}

We first examine how the density profile of the disk is affected by
the planet's radiative feedback. We compute the azimuthally-averaged
density profile $\langle \Sigma \rangle$, given by $\langle \Sigma \rangle = \frac{1}{2\pi}
\int_{0}^{2\pi} \Sigma(r,\phi) d\phi$.  In Figure \ref{SigmaProfile2}
we compare two models: the first one has no feedback ($L_{\rm p}=0$),
while the second one assumes a planet luminosity of $L_{\rm p}=10^{-3}
\, \rm L_\odot$.  The density profile are shown at 300 planet
orbits. At this stage, the gap carved by the planet is already formed
and in a quasi steady state.

For this model, the effects of planet feedback are noticeable in a
region of approximately 8~au radial extent about the planet.  We
notice that when the feedback is activated, the density profile
increases in the inner disk (the disk region inside the planet's
orbit), while the density decreases in the outer disk. Outside this
region, there is no apparent difference if the planet emits energy or
not. The observed changes in the density profile indicate that the
planet's luminosity enhances the ability of the disk to transport mass
from the outer to the inner disk through the protoplanetary
gap. Although not shown here, we have checked that this effect is
practically negligible for planet luminosities smaller than $10^{-3}
\, \rm L_\odot$.

\begin{figure} [h!]
\epsscale{0.8}
\plotone{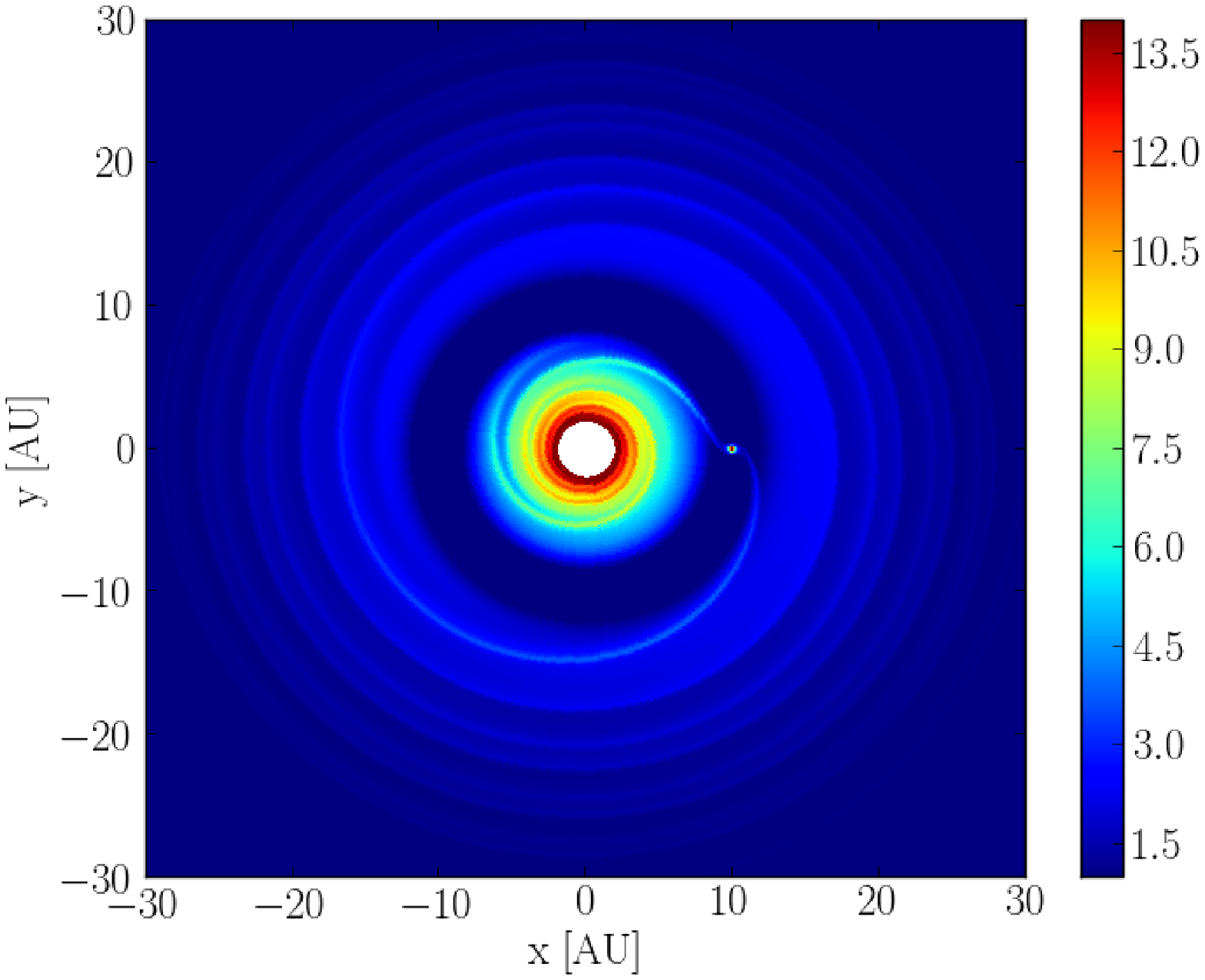}  
\plotone{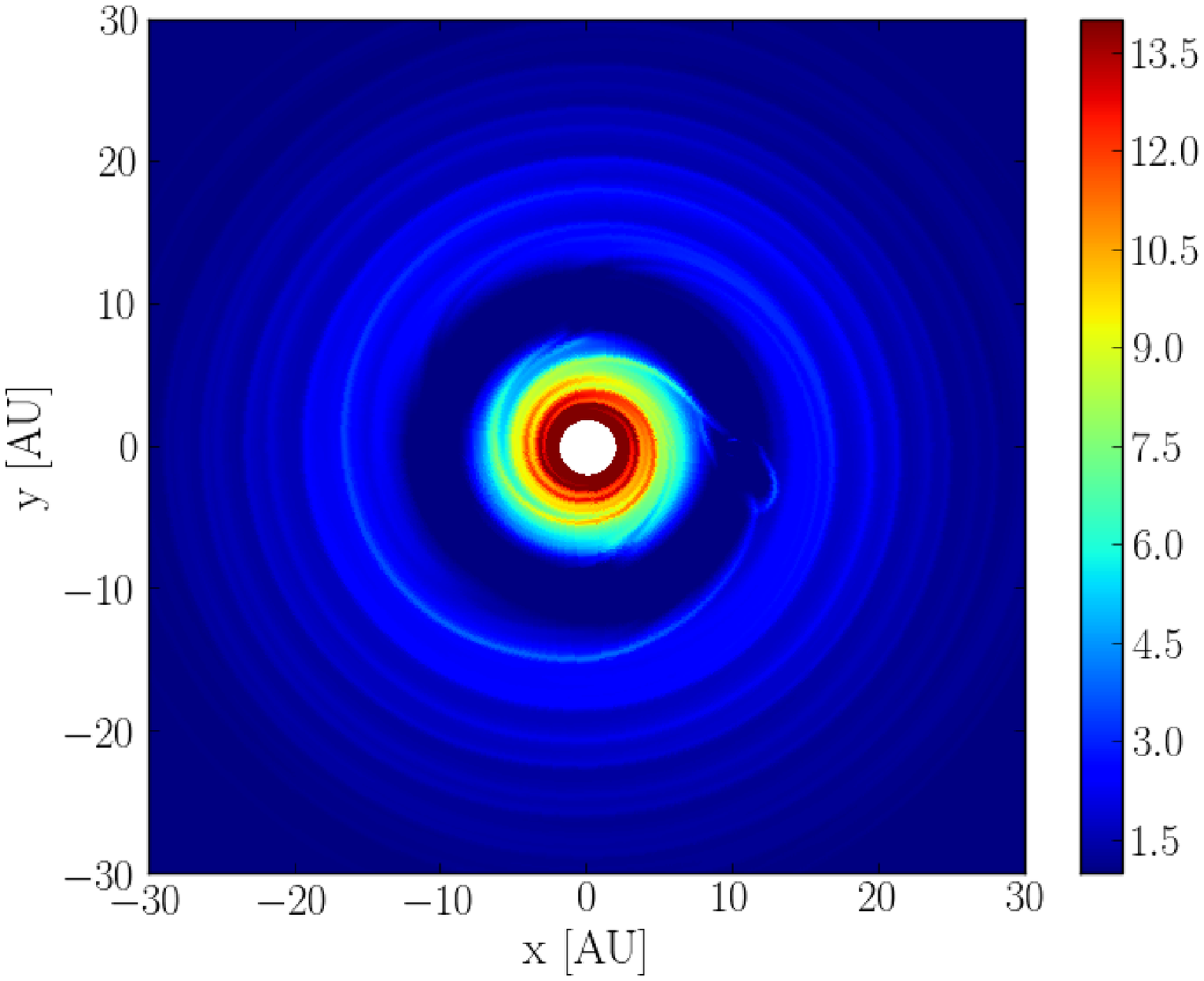}
\caption{Density contours (cgs units) of the disk at 300 orbits. The top panel
  shows our results without planet feedback. The bottom panel is for a
  model with $L_p = 10^{-3} \, \rm L_\odot$. When feedback is
  activated, the inner disk is denser (bottom panel) than without
  feedback. There is an enhancement of the transport of matter from
  the outer to the inner disk when feedback is activated. From the
  figure one can also note that there is no matter accumulating at the
  circumplanetary region.}
\label{Dens2D-comp}
\end{figure}

In Figure \ref{Dens2D-comp}, we compare density contours of the disk
with and without feedback after 300 planet orbits. As previously shown
by Figure~\ref{SigmaProfile2}, when feedback is activated the inner
disk becomes slightly denser, suggesting that the flux of matter from
the outer to the inner disk is enhanced by the planet feedback. 

It is worth mentioning that, when there is no feedback,  
 the increment of the inner disk density 
is much smaller than in the case the feedback is activated (Fig. \ref{Dens2D-comp}).
Recall that we use \textit{outflow} (inner/outer) boundary conditions, hence the density increment is not due to material accumulating at the boundary, but rather a result of a change in the stellocentric flux of matter stimulated near the planet (see \S~\ref{AccrRate}). Notice also that in our models there is no material accumulating at the circumplanetary
region. On the contrary, material is being slowly evacuated 
from the circumplanetary disk (as seen in Fig. \ref{Dens2D-comp}), but
after 1000 orbits there is still plenty of  material producing an optically thick
circumplanetary region.
Correspondingly, the density of the outer disk is slightly
decreases when the planet feedback is activated.  We will show later
that actually the disk's stellocentric accretion rate is enhanced at the planet location, explaining this behaviour.

\subsection{Temperature field}

\begin{figure} [h]
\plotone{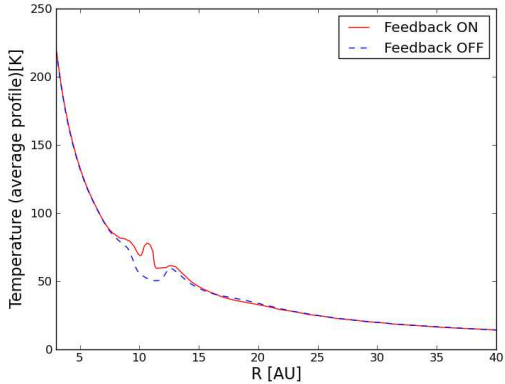}
\caption{Azimuthally-averaged temperature profile without and with
  feedback ($L_{\rm p}=10^{-3} \, \rm L_\odot$). Both profiles are
  taken after 300 orbits of the planet.}
\label{Temp1D}
\end{figure}

In Figure \ref{Temp1D} we show the azimuthally-averaged temperature
profile (i.e., $\langle T \rangle = \frac{1}{2\pi} \int_{0}^{2\pi} T(r,\phi) d\phi$)
of the disk for a model with a planet luminosity of $10^{-3} \, \rm
L_\odot$ and another without feedback, both after 300 orbits of the
planet. We see that at the planet's location the azimuthally-averaged
temperature increases from about 50 K without feedback to nearly 80 K
with feedback. Recall that these are  \textit{azimuthally-averaged}
temperature profiles, and do not reflect the large \textit{local} variation shown below.

\begin{figure} [h]
\epsscale{0.8}
\plotone{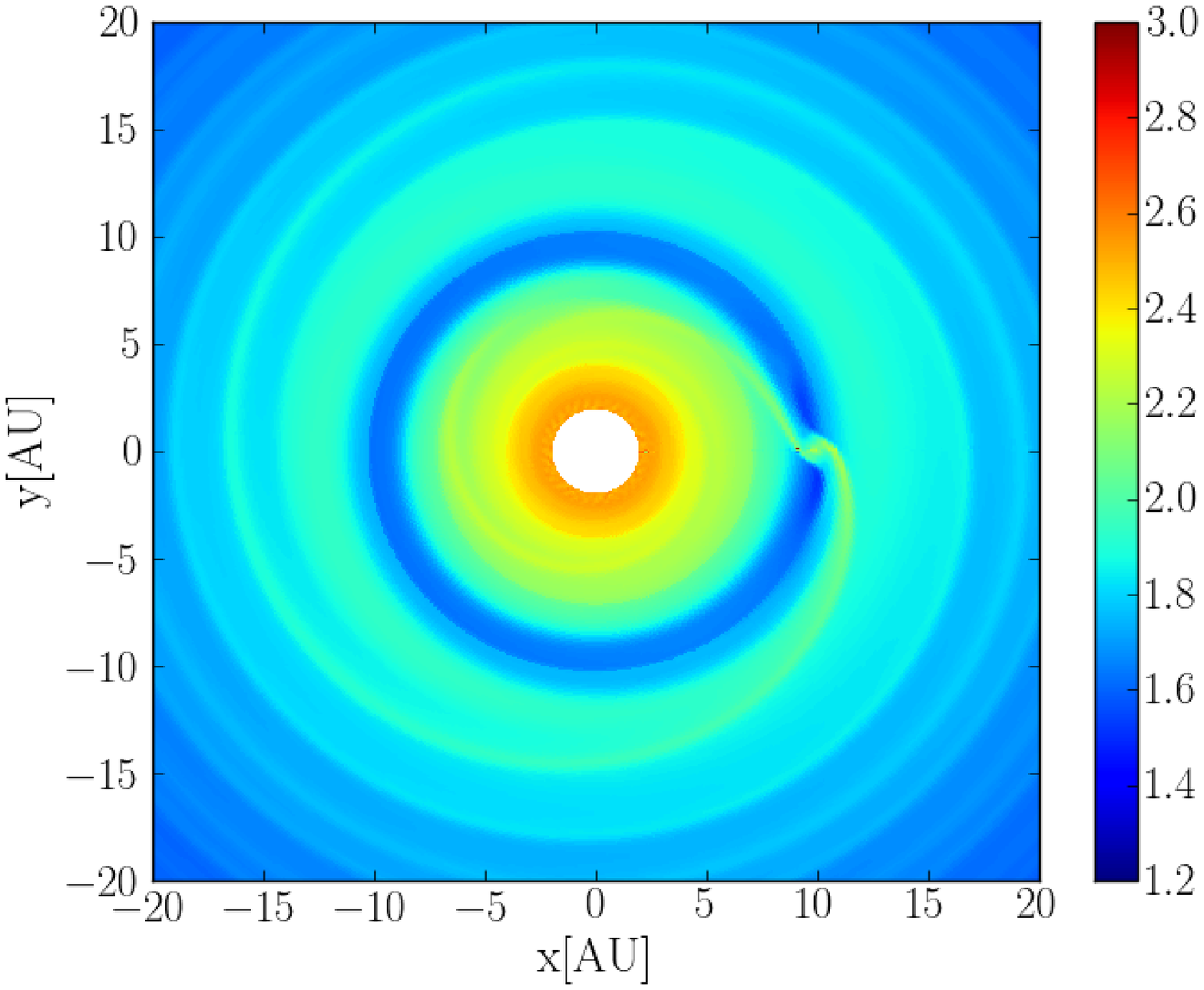}  
\plotone{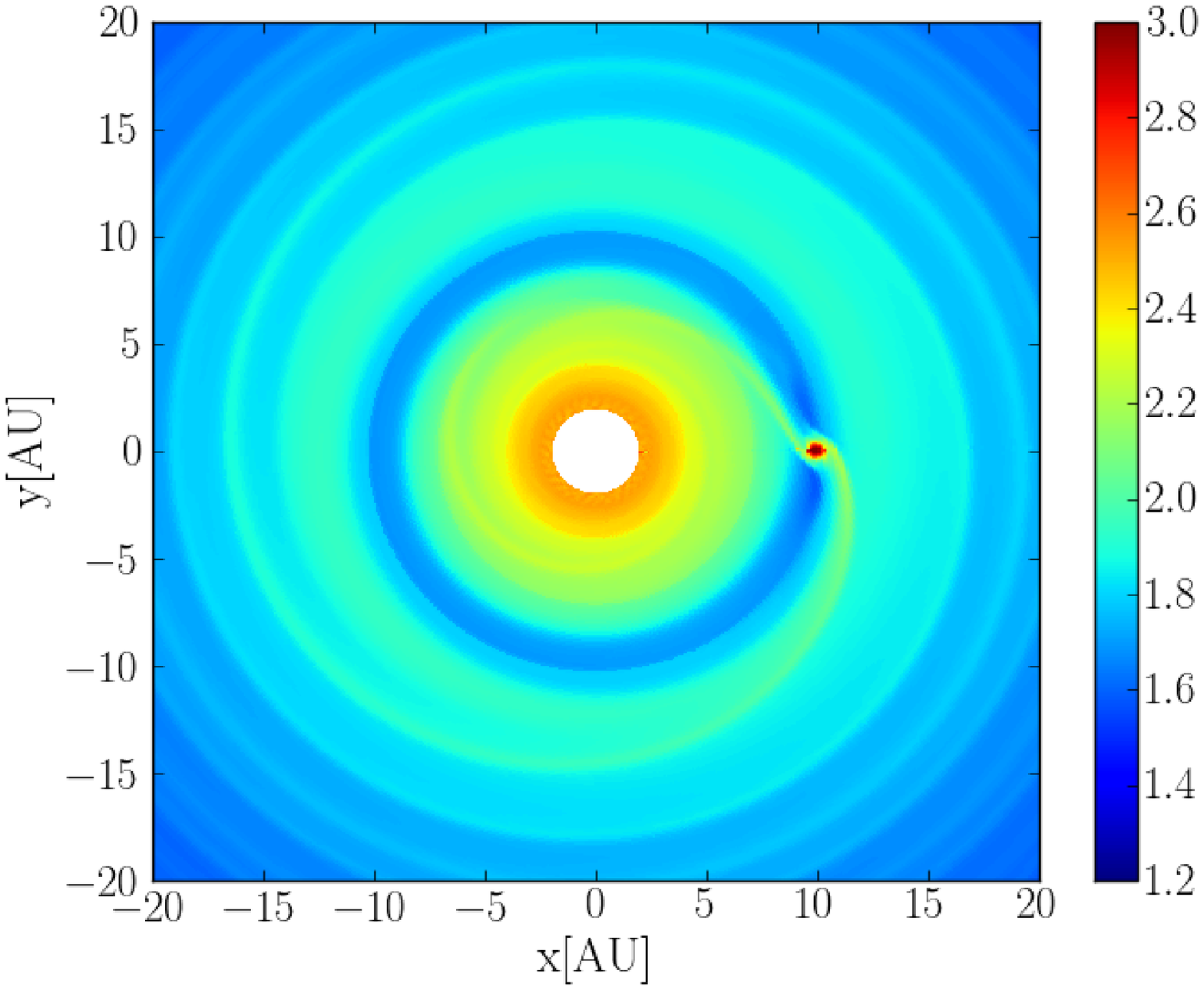}
\caption{Effective temperature of the disk after 300 orbits of the
  planet. The top panel shows our model without planet feedback, the
  bottom panel is for our model with $L_{\rm p}=10^{-3}\, \rm
  L_\odot$. Without feedback, the circumplanetary region reaches
  temperatures of $ 160 \, \rm K$, while when the planet luminosity is
  included this area reaches a peak temperature of about $ 1190 \, \rm
  K$.}
\label{Temp2D-comp}
\end{figure}

In Figure \ref{Temp2D-comp} we compare the surface
temperature of the disk after 300 orbits for the cases with $L_{\rm p}
= 0$ and $L_{\rm p}=10^{-3} \, \rm L_\odot$ (note that a logarithmic
scale is used, and that only the inner 20 au of the disk are shown).
When no feedback is included, the maximum temperature occurs at the
grid's innermost radius. The maximum temperature reached at the
planet's location is about $150 \, \rm K$. When the feedback is
activated, a hot spot forms within an au or so from the planet's
location, which is close to the size of the planet's circumplanetary
region (recall that in our feedback model, the energy released by the
planet is injected in a region of area $\pi R_{\rm CPD}^2$ about
the planet's location, where $R_{\rm CPD}$ denotes the radius of the
planet's circumplanetary material, which is $\sim 0.5$ au). The
maximum temperature reached in the planet's circumplanetary disk is
about $1190 \, \rm K$.

After 300 orbits of the planet, the aspect ratio at the planet
location is about $H_{\rm p}/r_{\rm p} \sim 0.2$, therefore the
pressure scale height of the disk at this position is $H_{\rm p} \sim
2 \, \rm au$. 

\subsection{Spectral signature}\label{SpecSignature}

\begin{figure} [h]
\plotone{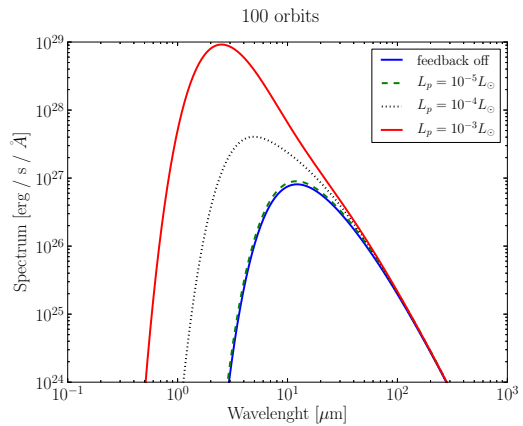}
\caption{Spectrum of the disk after 300 orbits without and with planet
  feedback, for $L_{\rm p}=10^{-5}$, $L_{\rm p}=10^{-4}$ and $L_{\rm
    p}=10^{-3}$ solar luminosities.}
\label{SpectrumAll}
\end{figure}

We have shown in the previous subsection that the disk temperature
near the planet's location is strongly enhanced by the inclusion of the
planet feedback. This enhancement (from $\sim 100 \, \rm K$ to $\sim
1000 \, \rm K$) should cause a significant variation in the spectral
emission of the disk in the vicinity of the planet. In
Fig.~\ref{SpectrumAll} we display after 300 orbits how the disk
spectrum, calculated from Eq. \ref{spec1}, changes for different
planet luminosities ($L_{\rm p}=10^{-5}$, $L_{\rm p}=10^{-4}$, $L_{\rm
  p}=10^{-3}$, and $L_{\rm p}=0$ $\rm L_\odot$). We see that the
spectrum is dramatically modified when the feedback is higher than
$L_{\rm p} \gtrsim 10^{-4} \, \rm L_\odot$.

At the planet's location the disk temperature without feedback is
about $150 \rm K$. When the feedback is included, the temperature
derived from the energy equation in the vicinity of the planet peaks
at $ 1100\,$K\footnote{We point out that the high temperatures we observe in our numerical models are also recovered by \cite{Zhu-2015}.   In his 1-dimensional accretion disk models the temperature down at the atmosphere of the planet ($r \sim R_{\rm J}$) reaches $\sim 2000\,$K, without invoking any energy input from the planet itself.  Notice that those scales cannot be resolved with our numerical scheme, so a direct comparison is not possible.} for planet luminosities $L_{\rm p} \gtrsim 10^{-3}
L_\odot$. This increase in temperature produces a corresponding
increase in the peak flux density by a factor of $\sim 100$, as seen
in Fig.~\ref{SpectrumAll}. Notice also that without feedback the
spectrum of the disk peaks at $19.3 \rm \mu m$, and that for a planet
luminosity of $L_{\rm p} = 10^{-3} \, \rm L_\odot$, the spectrum peaks
at $2.5 \rm \mu m$.

Using Eq.~\ref{Bolometric}, we can integrate the spectrum from
Fig.~\ref{SpectrumAll} to obtain the bolometric luminosity of the
entire disk\footnote{Notice that this bolometric luminosity does not
  include the emission from the star.  Moreover, the quoted values
  depend on the radial extent of the disk, most importantly on its
  inner radius, which we set for numerical convenience to 1 au.}.  In
the simulation without feedback, the bolometric luminosity is $L_{\rm
  b} = 5.54 \times 10^{-2} \, \rm L_\odot$, while in the case when we
assume a planet luminosity of $L_{\rm p} = 10^{-3} \, \rm L_\odot$,
the bolometric luminosity increases to $L_{\rm b} = 0.98 \rm
L_\odot$. In other words, the radiative feedback from the planet
results in a disk luminosity increased by a factor of 17.8.

\begin{figure} [h]
\plotone{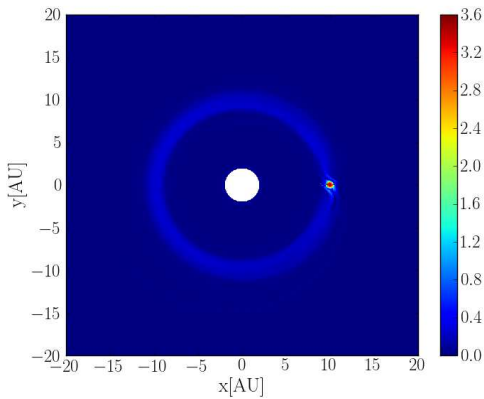}
\caption{Map ratio ($\log_{10}$ scale) of the bolometric emission per
  unit area between our model with $L_{\rm p}=10^{-3}\, \rm L_\odot$
  and our model with $L_{\rm p}=0$, at 300 orbits. When the feedback
  is activated, the bolometric surface brightness is about $10^{3.6}$
  times higher in the circumplanetary region, and there is also an
  excess of emission along the planet's orbit.}
\label{HD1-Radiation_ratio}
\end{figure}

To better appreciate the differences between the cases $L_{\rm p}=0$
and $L_{\rm p}=10^{-3} \, \rm L_\odot$, we plot in
Fig.~\ref{HD1-Radiation_ratio} (log scale) the ratio between the
bolometric emission per unit area (given by $\int_0^\infty
\Phi(T(r,\phi))\rm d\lambda = \sigma T^4$) for $L_{\rm p}=10^{-3}\,
\rm L_\odot$ and the emission when $L_{\rm p}=0$, after 300 orbits of
the planet.  From Fig.~\ref{HD1-Radiation_ratio} we notice that when
the feedback is activated, the blackbody emissions ($\sigma T^4$) at
the vicinity of the planet, inside a radius of about $\sim 5 \, \rm
au$, increases by up to 3.6 orders of magnitude over a situation
without feedback. Also, we notice that along the orbital path of the
planet around the central star, the feedback leaves a `track' of
enhanced disk surface brightness by about one order of magnitude. This
track indicates that the radiative feedback from the luminosity of the
planet induces net heating of the gas not only in the close vicinity
of the planet, but also along the planet's trajectory.

\begin{figure} [h]
\epsscale{0.8}
\plotone{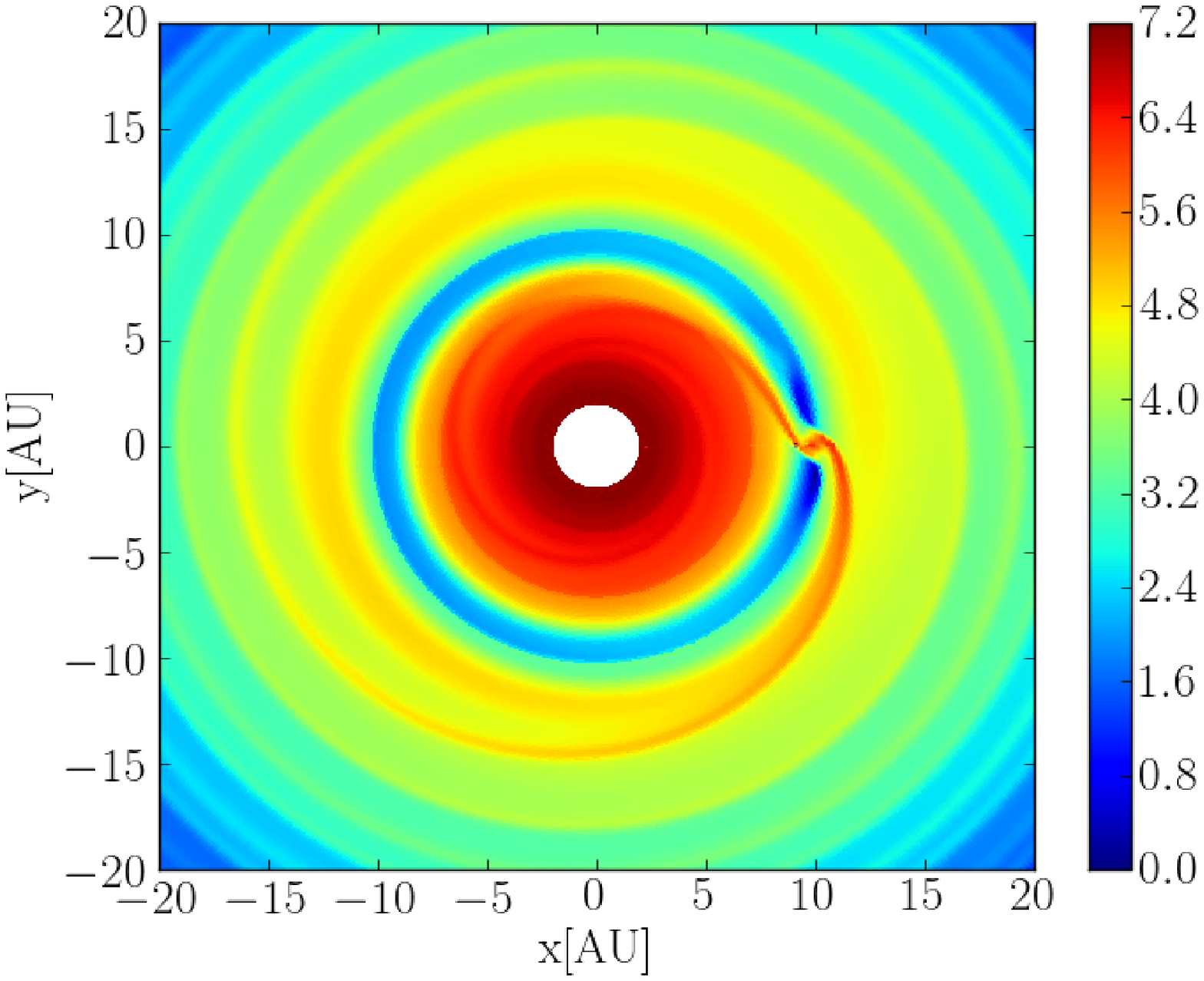}
\plotone{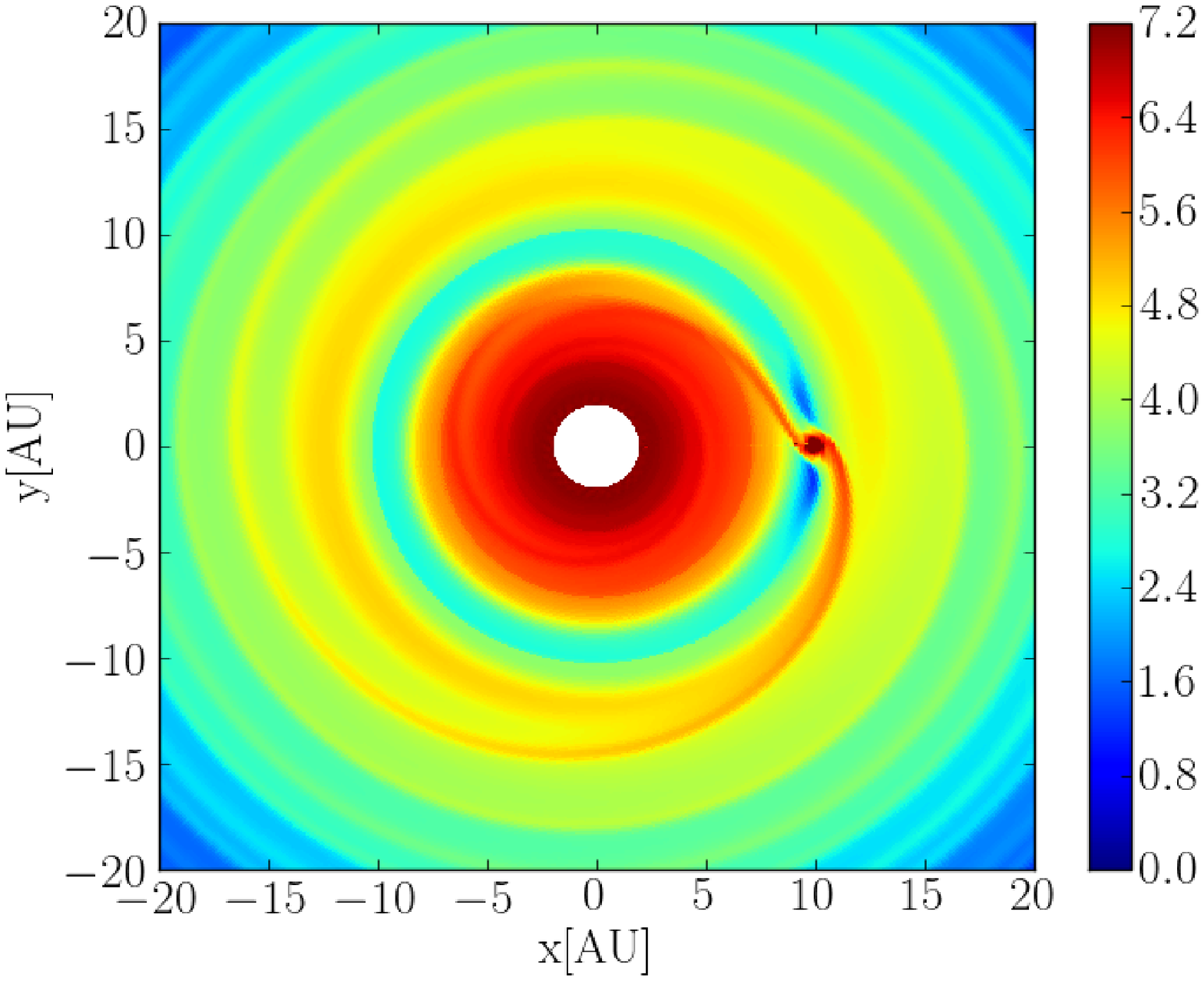}
\caption{Local blackbody flux per unit area $B_\lambda(T(r,\phi))$ of
  the disk according to Eq. \ref{Phi_nu}, taking $\lambda = 2.5 \mu
  m$.  The map is in log scale and the units are $\rm erg ~
  s^{-1}cm^{-3}$.  The upper panel corresponds to a model without
  feedback, the bottom panel is for $L_p = 10^{-3} \rm L_\odot$.  The
  highest surface brightness comes from the planet's circumplanetary
  region.}
\label{Flux2D-comp}
\end{figure}

In Figure \ref{Flux2D-comp}, we display at 300 planet orbits the
blackbody radiation $B_\lambda (T(r,\phi))$ for $\lambda = 2.5 \rm \mu
m$ for the cases without and with feedback ($L_{\rm p} = 10^{-3} \rm
L_\odot$). The feedback dramatically increases the blackbody radiation
in the circumplanetary region.  We point out that there could appear
to be an energy conservation problem here, because when there is no
feedback we get a luminosity output from the entire disk of about
$L_{\rm b} = 5.5 \times 10^{-2} \, \rm L_\odot$, while adding a
relatively small source of $L_{\rm p} = 10^{-3} \rm L_\odot$, we
obtain an output luminosity of $ 0.98 L_\odot$. In the next
subsection, we run several tests in order to check this result,
concluding that there is no inconsistency with energy
conservation. The additional energy comes from an increase on the
stellocentric accretion rate through the disk when the feedback is activated, as
will be explained in \S\ref{AccrRate} below.

\subsection{Feedback behaviour for other viscosity prescriptions}
\label{viscosities}

In this section we examine to what extent the impact of the planet's
radiative feedback depends on the assumption on the viscosity
prescription. For this purpose, we have carried out a number of
simulations with different viscosity prescriptions.

We first employed alpha viscous disk models. The alpha disk model, as
it is implemented in the public release of FARGO-AD, assumes a
kinematic viscosity $\nu = \alpha \langle c^2_{\rm s} / \Omega \rangle$ where the
brackets stand for the azimuthal average. Differently said, the
kinematic viscosity is purely radial. We have carried out simulations
with $\alpha = 4\times 10^{-4}$, $4\times 10^{-3}$ and $4\times
10^{-2}$, and found nearly identical peak temperature in the planet's
circumplanetary region. Our results are listed in Table \ref{Table1},
along with disk's accretion rates at the planet location, which we will
detail in the next section.

Lastly, we ran an inviscid model, i.e., a model without shear
viscosity ($\nu=0$; note, however, that this model still includes
artificial viscous heating via a bulk viscosity, as in all the models
presented in this paper -- see model described in
\S~\ref{sec:viscousdiss}). In that case, we find again a peak
temperature near the planet of $1166 \, \rm K$, which is very similar
to the cases presented before with different viscosity
prescriptions. All these numerical experiments show that the peak
temperature that we find with planet feedback activated is basically
independent of the disk's viscosity.

\begin{table}[h!]
\small
\centering
\scalebox{0.8}{
\begin{tabular}{| l||r|r|r|c|}  
  \hline
  \textbf{Viscosity model} & \textbf{$T_{peak}^{\rm on}$} [K] & \textbf{$T_{peak}^{\rm off}$} [K] & $\dot{M}_{\rm on} [\rm M_\odot/yr]$ & $\dot{M}_{\rm off} [\rm M_\odot/yr]$\\
  \hline\hline
  $\alpha = 4 \times 10^{-4}$ & $ 1170$ & $ 125$  &   $3.7 \times 10^{-6}$& $2 \times 10^{-8}$  \\  
  \hline
  $\alpha = 4 \times 10^{-3}$ & $ 1190$  & $ 200$ & $2.6 \times 10^{-6}$ & $1.2 \times 10^{-7}$ \\
  \hline
  $\alpha = 4 \times 10^{-2}$ & $ 1200$ &  $ 276$ & $3.8 \times 10^{-6}$ &  $2.3 \times 10^{-7}$ \\
  \hline
  $\nu = 0$ & $ 1166$ & $ 100$ & $1 \times 10^{-6}$ & $1 \times 10^{-8}$  \\
  \hline
\end{tabular}}
\caption{Peak temperature and disk accretion rate values at the circumplanetary  region for different viscosity prescriptions, with feedback (columns 1 and 3 on 
  the right-hand side, $L_{\rm p} = 10^{-3} \, \rm L_\odot$) and without 
  feedback (columns 2 and 4).}
\label{Table1}
\end{table}

\subsection{Accretion rate}\label{AccrRate}
\label{sec:accretionrate}
We have shown in \S~\ref{SpecSignature} that the bolometric luminosity
of the disk increases by a factor of $\sim 20$ with planet feedback
and a planet luminosity of $L_{\rm p}=10^{-3} \rm L_\odot$, compared
to a disk without planet feedback. This excess of energy mostly arises
from the planet's circumplanetary region and from a narrow region
about the planet's orbital radius (see Figure
\ref{HD1-Radiation_ratio}). The luminosity can be related to an
accretion rate, an excess of luminosity should therefore have a
corresponding increase in the disk's accretion rate.

\begin{figure} [h]
\epsscale{0.8}
\plotone{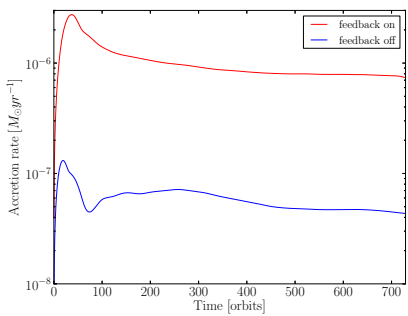}
\caption{Disk accretion rate at the planet's orbital radius, defined by
  Eq.~(\ref{eq:Mdot}), without and with ($L_{\rm p}=10^{-3} \, \rm
  L_\odot$) planet feedback.  With feedback the accretion rate at the
  planet's orbital radius is enhanced by an order of magnitude. The
  peak of the accretion rate is reached near $\sim 40$ planet orbits,
  at this stage the temperature reaches a maximum near the planet
  vicinity.}
\label{Acc_gap}
\end{figure}

We calculate the azimuthally-integrated accretion rate of the disk at the planet
location as
\begin{equation}
\dot{M} = \int  v_r(r_{\rm p},\phi) \Sigma(r_{\rm p},\phi) r_{\rm p} d\phi,
\label{eq:Mdot}
\end{equation} 
where $v_r$ is the radial velocity of the gas. In Figure~\ref{Acc_gap}
we display the time evolution of $\dot{M}$ for two models: one with
feedback ($L_{\rm p}=10^{-3}$) and one without feedback ($L_{\rm
  p}=0$), assuming an alpha disk with $\alpha = 4 \times 10^{-3}$ (our
fiducial model). We see that, when the feedback is activated, the
disk's accretion rate is enhanced by an order of magnitude.  The accretion
rate reaches similar values for different viscosity models, see Table
\ref{Table1}.

\begin{figure} [h]
\epsscale{0.9}
\plotone{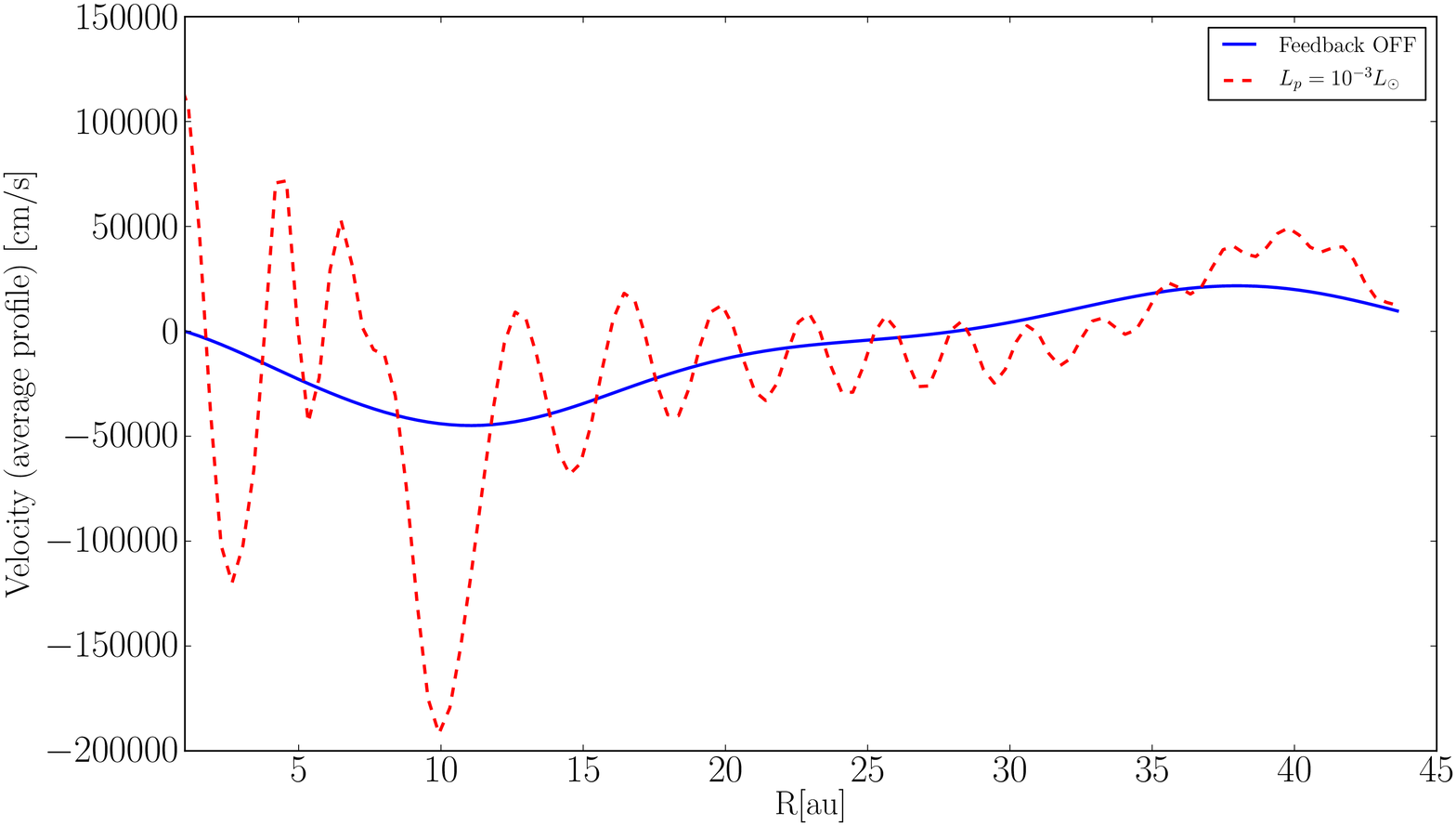} %
\caption{Azimuthal average of the radial velocity after 600 orbits.  When the feedback is activated, the radial velocity is enhanced by a factor $\sim 4$ at the planet location (10 AU).}
\label{Vrad-profile}
\end{figure}

We notice from Figure \ref{SigmaProfile2} that the density average at the planet location did not differ much between the cases with and without feedback, hence, in order  to have an enhancement of the accretion rate at that region, it must be the radial velocity that is enhanced. This is exactly what we see in Figure \ref{Vrad-profile}, where the radial velocity is enhanced by a factor $\sim 4$ when the feedback is activated. 
It is not expected to have a similar enhancement factor as the observed for the accretion rate mentioned above (i.e., one order of magnitude) because Figures \ref{SigmaProfile2} and \ref{Vrad-profile} show an \textit{azimuthally-average} radial velocity, rather than local values at the planet location. To compute $\dot{M}(r_p)$ as indicated in the precedent paragraph we use local values (where, for instance, the density takes  larger values at $r_p$ than the azimuthally averaged value), then we integrate over the azimuthal coordinate.

At the end of \S~\ref{SpecSignature}, we discuss the fact that our
simulations show, for instance, that when we use a local energy input
of $L_{\rm p} = 10^{-3} \, \rm L_\odot$, we obtain an energy output of
$L_{\rm b} \sim \rm L_\odot$. It could be misinterpreted as a
non-conservative evolution of the energy. To show that it is not the
case, we argue that the origin of the extra energy comes from an
increase in the disk's accretion rate stimulated by the feedback of the
planet, and that this enhancement is independent of the viscosity
prescription.

When radiation feedback is active, the luminosity of the disk at the location of the planet is proportional to the stellocentric disk accretion rate (which attains a maximum value), i.e., $L_{\rm acc} \propto \dot{M}$ at $R_p$. For instance, taking the model with $\alpha = 4 \times 10^{-3}$, the accretion rate at 300 orbits is $\dot{M}^{\rm on} \approx 10^{-6} \rm M_\odot yr^{-1}$ with feedback, and $\dot{M}^{\rm off} = 6 \times 10^{-8} \rm M_\odot yr^{-1}$ without (see Figure \ref{Acc_gap}). From this, we find that the accretion luminosity of the disk should be increased by a factor of roughly  $L_{\rm acc}^{\rm on}/L_{\rm acc}^{\rm off} = \dot{M}^{\rm on}/\dot{M}^{\rm off} \simeq 17$  by the inclusion of planet feedback. This is indeed in good agreement with the increase in bolometric luminosity calculated in \S~\ref{SpecSignature}, where the luminosity increase is found to be $\approx 17.8$. Our results clearly
show that there is no imbalance in the energy bill during the
simulation, and that the extra energy is the direct product of an
enhancement in the disk accretion rate, which in turn was stimulated by the
planet feedback.

\begin{figure} [h]
\epsscale{0.8}
\plotone{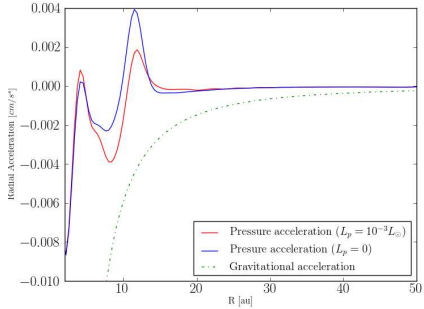} %
\caption{Azimuthal average of the radial acceleration due to the
  pressure work done by the disk without and with feedback. The
  gravitational radial acceleration is displayed for comparison.}
\label{gradP}
\end{figure}

Now we briefly describe the effect on the disk dynamics when the
radiative feedback of the planet is included. From a dynamical point
of view, the radial component of the Navier-Stokes equation includes
the radial pressure gradient of the fluid (the term
$(1/\Sigma) \partial P / \partial r$). A local enhancement of this
term at the planet location should lead to a local increase in the
radial velocity (understood in absolute value). In Figure \ref{gradP},
we compare the azimuthally-averaged radial acceleration due to the
gradient pressure when the feedback is turned on ($L_p = 10^{-3} \rm
L_\odot$) and off, after 100 planet orbits. We show for comparison the
radial gravitational acceleration due to the central star. We see that
the pressure gradient is particularly strong near the planet's
location, as expected. Moreover, with feedback the radial acceleration
is more negative, which acts to increase the accretion rate through
the disk near the planet's location, as seen before. 

Similar conclusions were obtained by \cite{Owen-2014}, who construct a 1D
radial 'transition' disk model including feedback from the protoplanet
acting over the gas and dust components of the disk. They include the
radial and azimuthal accelerations of the dust particles produced by
radiative feedback from the planet in a secular model of a
protoplanetary disk with an embedded accreting planet (e.g., \cite{Clarke-Pringle-1988}), finding that the accretion rates observed in transition
disks are better explained when the radiative feedback of the planet
is included. The \cite{Owen-2014} model assumes azimuthal symmetry, in
which the radiation feedback and the disk-planet interaction are
treated as average quantities. By contrast, our model is
non-axisymmetric (2D), and we include a non-stationary energy equation
to introduce the radiative feedback.

It is important to clarify that this accretion enhancement 
should lead to a transient situation explained as this: as shown above, 
the local feedback promotes the flux of matter in a region close to the circumplanetary disk.
As seen in Figure \ref{Dens2D-comp}, gas density slowly diminishes at the CPD. If this continues long enough, the CPD will be depleted becoming optically thin.
 If this happens, radiation from the planet will no longer interact with the gas, escaping immediately  from the disk without heating it. Therefore, the radiative feedback will no longer affect the CPD, muting its effect on the gas dynamics. Consequently, accretion enhancement will be no more, as well as the produced extra luminosity. Once the CPD start again
to accumulate material  and became anew optically thick, radiation feedback
will interact one more time with the gas, modifying its dynamics as before. From our calculations, to deplete the CPD with a planet feedback of $\sim 10^{-3} ~ \rm L_\odot$ located at 10 au, takes more than 10 thousand years, but further investigations are needed to clarify this point.

To summarize, our findings show that a relatively small source of
heating coming from the planet's circumplanetary region will induce
enhanced disk accretion rates near or around the planet's orbital
radius. This enhancement facilitates the extraction of gravitational
energy, which locally heats the gas, enhancing the luminosity in the
circumplanetary region. This local increase in the disk temperature
then leads to an increase in the accretion rate, increasing even more
the temperature (and thus the luminosity). This effect results in a
positive feedback until a thermal balance is reached where the
dissipation heat rate equals the radiative cooling rate ($Q^+ = Q^-$
with our notations).

These accretion luminosities should be detectable inside gaps. It
should be noticed that it is not the intrinsic luminosity of the
planet, but a local gas luminosity stimulated by the action of the
planet feedback. In the next subsection, we present results of
hydrodynamical simulations dedicated to interpret the observations of
a protoplanet candidate around star HD 100546.

\subsection{HD~100546 simulation}
\label{sec:HD100546}
A protoplanet injecting extra local heating in the disk via the
feedback mechanism explained in the previous sections could explain
the bright compact emission detected in $L^\prime$ band with NACO in
the disk of HD 100546 \citep{Quanz-et-al-2013b}. HD 100546 is a Herbig
Ae/Be star which harbours a protoplanet candidate orbiting at 68~au
(\cite{Currie-et-al-2014}, \cite{Quanz-et-al-2013a}). Motivated by these
observations, we have carried out a set of simulations for HD
100546. We compare the results of various accretion luminosities with
the luminosity of the observed planet candidate.

Based on interferometric data using AMBER/VLTI and photometric
observations, \cite{Tatulli-et-al-2011} proposed a disk model for the
circumstellar environment of HD 100546. Using this model we adopt a
disk scale height profile $H(r) = 12\, (r /100 \, \rm au)^{1.1} \, \rm
au$, and a surface density profile $\Sigma(r) \propto r^{-1}$ as
initial conditions for our simulations. We tailor the disk mass in
order to guarantee that the circumplanetary region remains optically
thick. We thus set the disk gas mass to $M_d = 15 \times 10^{-2} \rm
M_\odot$ (while \cite{Tatulli-et-al-2011} reported $5 \times 10^{-2} \rm
M_\odot$ for gas). From \cite{Panic-et-al-2010}, we assume a central
star of mass $M_{\star} = 2.5 \rm M_\odot$ (which defines our unit of
mass), implying a planet's orbital period of 354.6 yr.  Neither
irradiation nor flux diffusion from the central star are taken into
account in our simulation.  In \S~\ref{Irradiation}, we justify this
approximation for the irradiation through a simple analytical
calculation. As in our fiducial model, the disk is treated as an alpha
viscous disk with $\alpha = 4 \times 10^{-3}$.

The planet is located at $r_{\rm p} = 68 \, \rm au$, which is taken as
the code's unit of length. For numerical convenience, the grid now
extends from 10 to 200 au along the radial direction, which
corresponds to radii ranging from 0.15 to 3.0 in code units.

Informed by our dust model for HD 100546 (see Sec.~\ref{RadT} below),
we changed the opacity prescription in our feedback calculation taking
into account an average opacity for a mix of dust species given by
$\kappa = 130 \, \rm cm^2/g$.  This is more adequate for transition
disks such as HD 100546.

We carried out a set of simulations with various planet luminosities:
$L_{\rm p} = \{0; \, 2.5; \, 5\} \times 10^{-4}\, \rm L_\odot$. The
observed emission in $L^\prime$ band is thought to correspond to a
protoplanet with a mass in between 1 and 8 $M_J$ \citep{Quanz-et-al-2013b}.
We fix the planet mass to $M_{\rm p} = 5 M_{\rm J}$.  Our results of
simulations are shown after 600 planet orbits, or $ 24.2 \times 10^3
\,\rm yr$ (as in the previous sections, the planet's mass reaches its
imposed value over 10 orbital periods).

\begin{figure}[h!]
  \includegraphics[scale=.3]{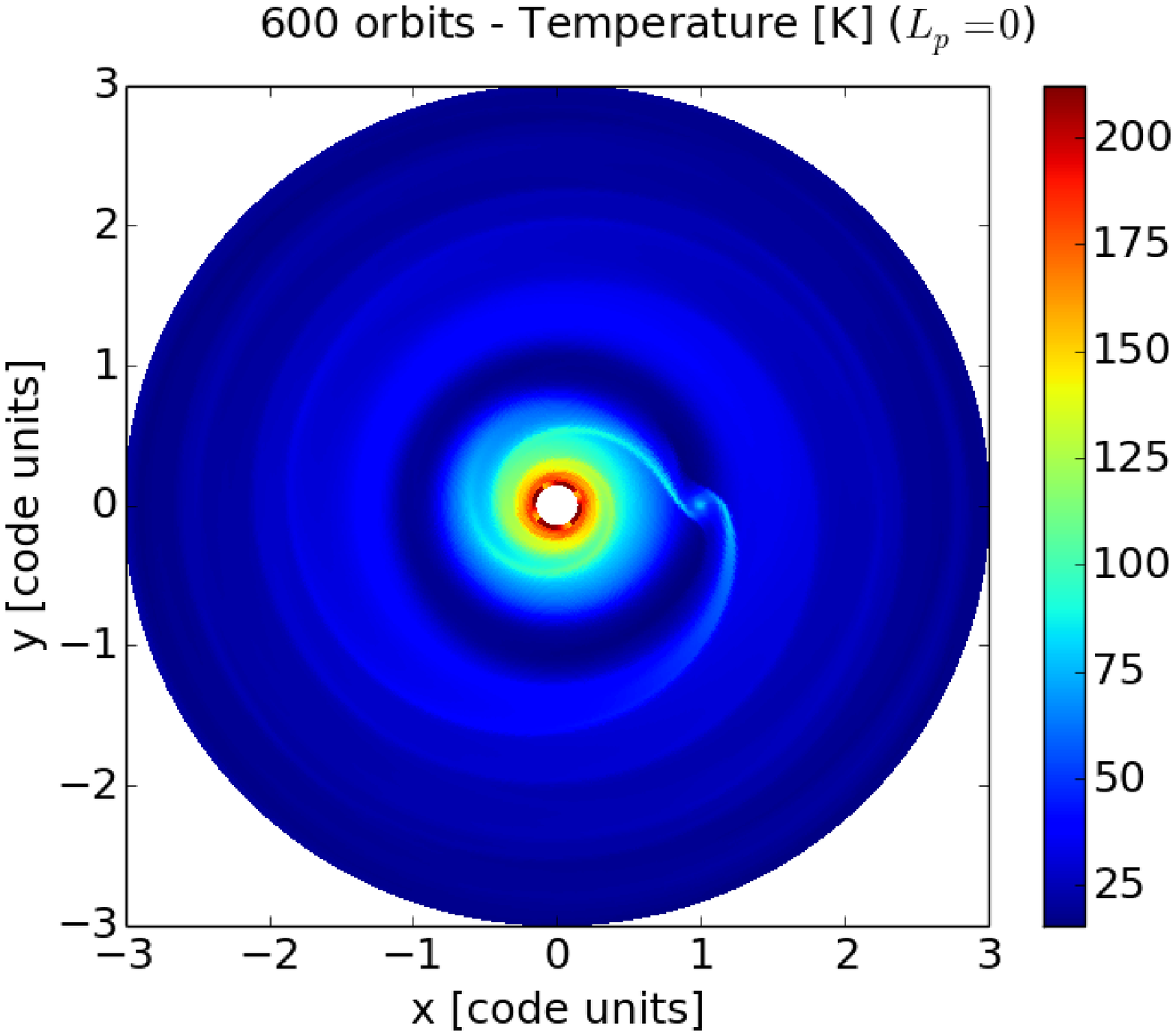}
  \includegraphics[scale=.3]{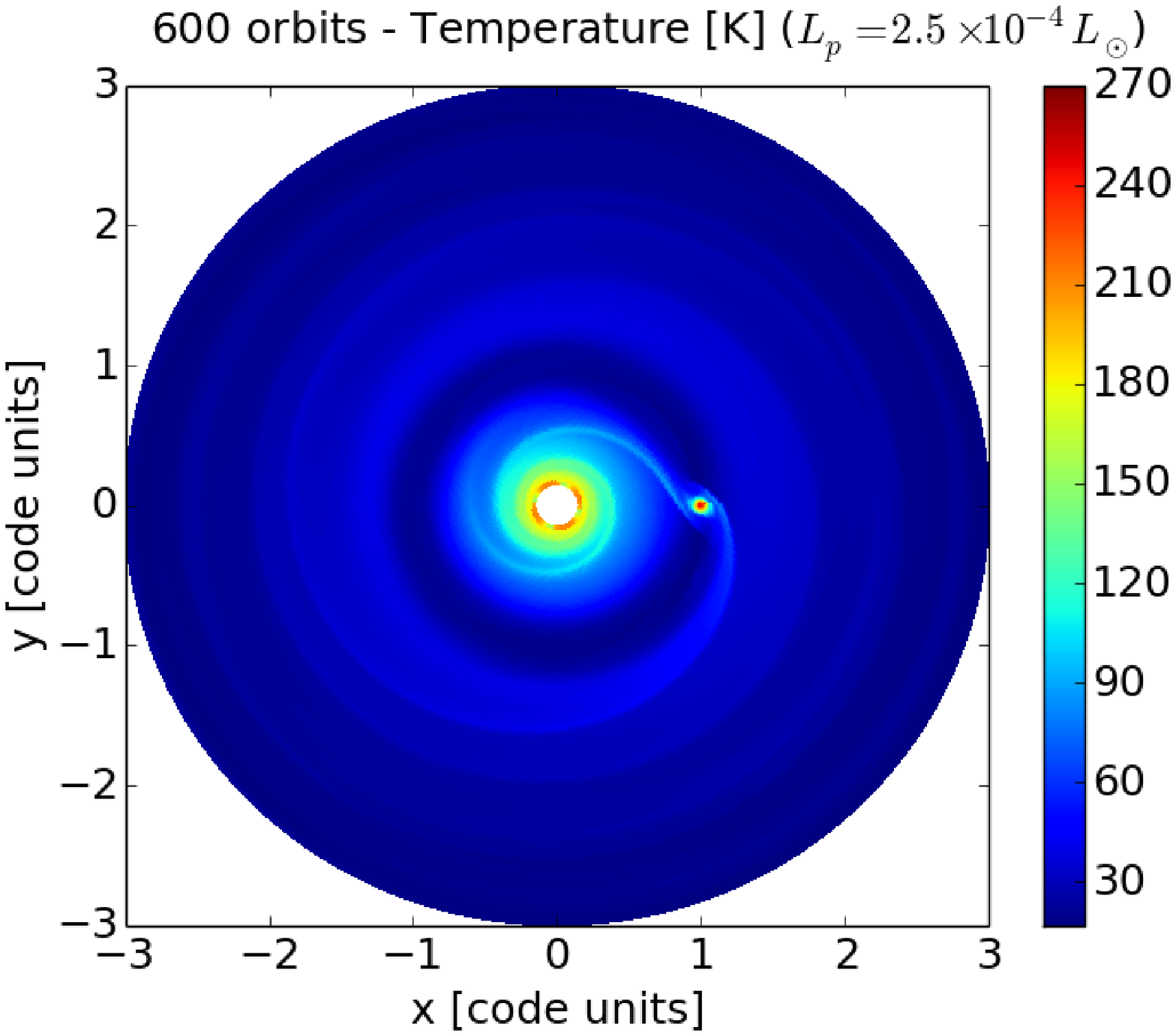}
  \includegraphics[scale=.3]{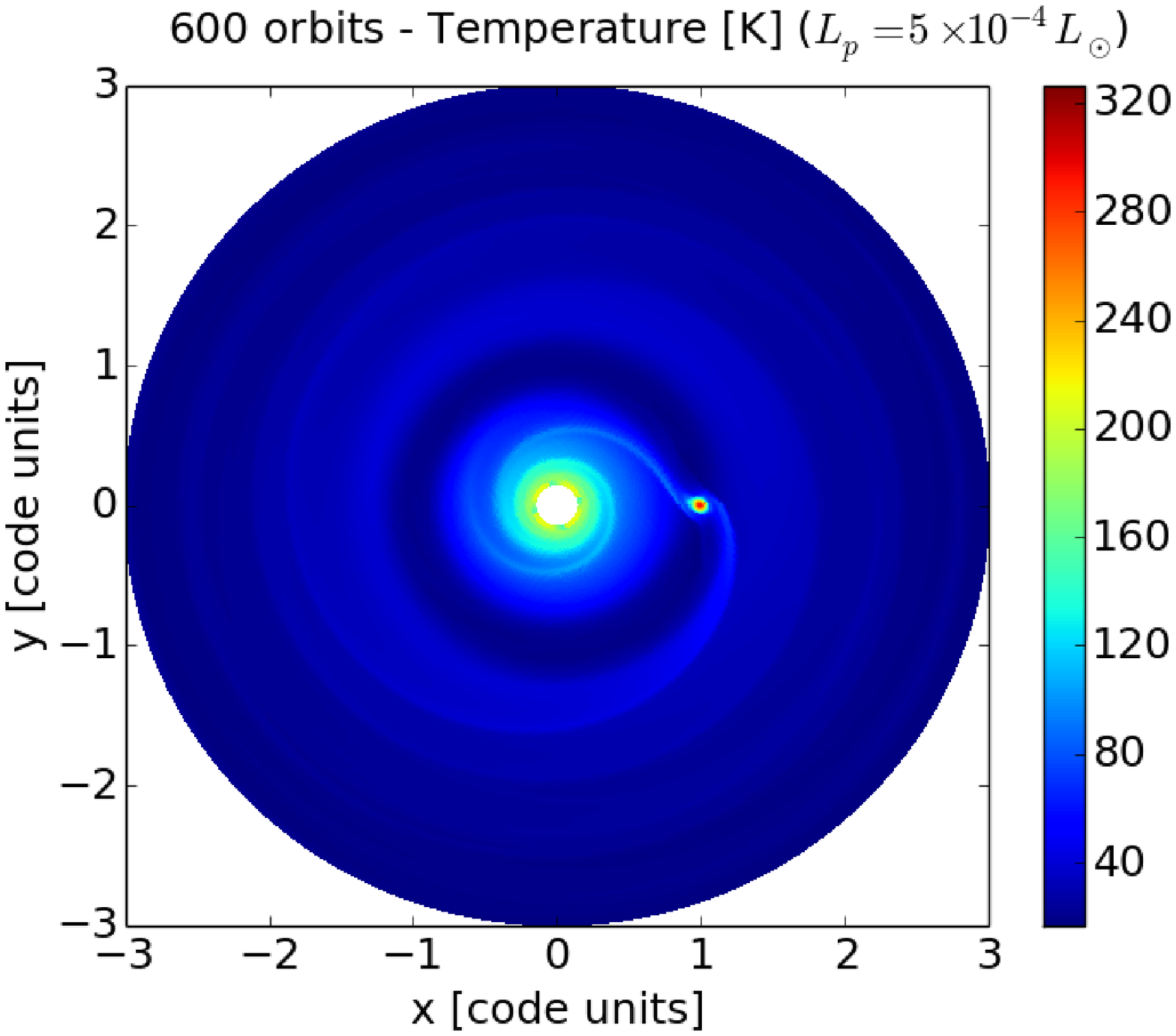}
  \caption{Contours of the disk temperature after 600 planet
    orbits. From top to bottom: $L_{\rm p} = 0$, $L_{\rm p} = 2.5
    \times 10^{-4} \, \rm L_\odot$, and $L_{\rm p} = 5 \times 10^{-4}
    L_\odot$. The peak temperature near the planet is 80 K, 270 K and
    325 K, respectively.}
  \label{HDTemp}
\end{figure}

We show in Figure \ref{HDTemp} contours of the disk temperature for
the different planet luminosities mentioned above.  The coordinates in
the figures are in code units, therefore the planet (at 68 au) appears
at $r = 1$.

The top panel corresponds to a simulation without feedback. In this
case, the hottest temperatures come from the innermost region of the
disk ($200 \, \rm K$). The circumplanetary region reaches a maximum
temperature of about $80 \, \rm K$. The middle panel shows a model
with feedback and a planet luminosity $L_{\rm p} = 2.5 \times 10^{-4}
\, \rm L_\odot$. In this case, the proto-planet region peaks at $270
\rm K$. In the lower panel, $L_{\rm p} = 5 \times 10^{-4} \, \rm
L_\odot$, which gives a peak temperature (at the planet location) of
$325 \rm K$.

By integrating the spectrum over the whole disk surface and over the
entire wavelength range (Eq. \ref{Bolometric}), we obtain the
bolometric luminosity of the disk.  Without feedback, it yields
$L_{\rm b}^{\rm off} = 4.2 \, \rm L_\odot$.  For $L_{\rm p} = 2.5
\times 10^{-4} \, \rm L_\odot$, we obtain a disk luminosity of about
$L_{\rm b} = 4.7 \, \rm L_\odot$, resulting in a disk 1.1 times
brighter than without feedback (i.e., $L_{\rm on}/L_{\rm off} =
1.1$). This can be explained by the fact that the addition of a small
source of heating near the planet (associated with $L_p = 2.5 \times
10^{-4} \, \rm L_\odot$) locally enhances the disk accretion rate at the
planet location, and results in a positive feedback able to produce a
disk that is $\sim 10\%$ brighter than without feedback. We point out
that the disk luminosity without feedback is 4.3 solar luminosities
(for a disk extending from 10 to 200 au), which is consistent with the
bolometric luminosity of $\sim$10~$L_\odot$ reported by Benisty et
al. (2010) (for a disk from 10 to 500 au). Recall that we neglect
irradiation from the star, and that modelling a smaller than observed
disk is likely to explain the factor $\sim 2$ discrepancy.

\subsubsection{Radiative transfer}\label{RadT}

In order to compare our HD~100546 simulation with the $L^\prime$
observations from \cite{Quanz-et-al-2013a}, we input the hydrodynamical
gas density and gas temperature fields into the {\sc
  RADMC3D}\footnote{http://www.ita.uni-heidelberg.de/$\rm \sim$dullemond/software/radmc-3d/} radiative transfer code (version 0.38,
  \citep{RADMC3D0.38}).  At infrared wavelengths the emission is
dominated by thermal and scattered emission from micron-sized dust.
Small grains should be well mixed, hence we assume that the dust
distribution follows the same density field as the gas in the
simulation. These grains should account for the bulk of the $L^\prime$
emission.

Our model dust distribution consists of a mix of 2 common species:
amorphous carbons and astronomical silicates \citep{Draine-Lee-1984}. As
informed by previous SED modeling (\cite{Benisty-et-al-2010}, \cite{Tatulli-et-al-2011}), our grain size distribution follows a power-law with
exponent $-3.5$, and particle sizes ranging from 0.05 to 1000~$\mu$m.
We used Mie theory (homogeneous spheres) to compute the dust opacities
for anisotropic scattering.  The optical constants were taken from \cite{Li-Greenberg-1997} for amorphous carbon grains, and from \cite{Draine-Lee-1984} for silicates. The intrinsic densities of the grains are
2~g~cm$^{-3}$ and 4~g~cm$^{-3}$ for amorphous carbons and silicates,
respectively. We also assume $T_{\rm gas} = T_{\rm dust}$.

\begin{figure*}[t]
\centering \includegraphics[width=0.65\textwidth]{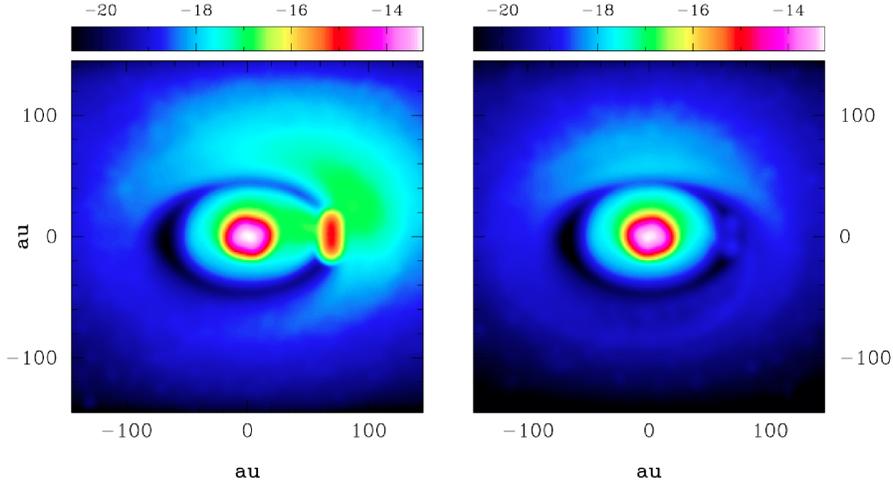}
\caption{Radiative transfer predictions in $L^\prime$ band
  (3.8~$\mu$m), for a disk inclination angle of 47 deg, calculated on
  the density and temperature fields shown in Fig. \ref{HDTemp}. The
  left panel shows the results with radiative feedback ($L_{\rm p} =
  2.5 \times 10^{-4} \, \rm L_\odot$), the right panel shows the case
  without feedback. The radiative feedback results in a
  circumplanetary hotspot reaching $\sim$1 mJy in $L^\prime$, close to
  the 1.3 mJy observed by \cite{Quanz-et-al-2013a}. The inclusion of the
  inner disk (with radius $<$10~au) would result in a bright stellar
  point source, which could efficiently be cancelled by Angular
  Differential Imaging (ADI). The images have been convolved with a
  0.1 arcsec Gaussian PSF.  Colour stretch is logarithmic.  Images are
  in units of $\rm \, erg \, cm^{-2} \, s^{-1} \, str^{-1}$.}
\label{radmc}
\end{figure*}

The 2D surface densities for each dust species are extended vertically
following hydrostatic equilibrium. This produces a 3D volume with a
Gaussian profile in the vertical direction. The disk scale-height is
assumed to be the same for both dust species and it is set by the
temperature field through the relation $H(r)=\sqrt{\gamma\,
  \overline{R} \,T(r)}\,r/v_{\rm Kep}$, where $v_{\rm Kep}$ is the
Keplerian velocity. For simplicity the temperature was assumed constant
in the vertical direction. A better vertical description of the
temperature would require a full account of stellar irradiation and
stellar flux diffusion in the simulations.  The final image is
produced performing a second order volume ray-tracing method. The
results of our radiative transfer calculation for $L^\prime$ are shown
in Fig.~\ref{radmc}.

\cite{Quanz-et-al-2013a} calculated the luminosity of the hot spot around
the companion candidate HD~100546b, assuming a point source embedded
in a region 0.1 arcsec in size ($\sim 10 \, \rm au$) located at 68 au
from the central star, reporting an apparent $L^\prime$ magnitude of
13.2, which translates into a flux density of 1.3 mJy. In order to
compare with our planet feedback predictions, we computed the emergent
flux from the vicinity of the protoplanet in the same way.

The simulation including an embedded protoplanet with an accretion
luminosity of $L_{\rm p} = 2.5 \times 10^{-4} \, \rm L_\odot$ yields
an emerging flux from the circumplanetary vicinity of 0.7~mJy in
$L^\prime$.  On the other hand, an accretion luminosity of $L_p = 5
\times 10^{-4} \, L_\odot$ already produces a flux density of
$\sim$3~Jy, about two times higher than the flux levels reported by
\cite{Quanz-et-al-2013a}.

A key result from our model is that the emitted luminosity from the
protoplanet region depends on the prescribed feedback and not on the
viscosity prescription. Therefore, from our simulations for
HD~100546b, we conclude that the proto-planet candidate is compatible
with an accreting 5 mass Jupiter-like object, with a planet luminosity
of about $\sim 2.5 \times 10^{-4}\, \rm L_\odot$.

\section{Discussion}\label{discussion}

\subsection{Irradiation from the central star}\label{Irradiation}

Here we briefly discuss the possible effect of the irradiation from
the central star on the disk.
The flux penetrating the surface of the disk can be
calculated as (see for instance \cite{Frohlich-2003})

\begin{equation}
F_{\rm irr} = (1 - \beta) \frac{L_{\star}}{4 \pi r^2} \cos{\varphi}, 
\end{equation}
where $\beta$ is the albedo (reflection coefficient), and $\varphi$ is
the angle formed by the incident radiation and the normal to the
surface. It can be shown that $\cos{\varphi} \simeq dH/dr - H/r$, and
using $F_{\rm irr} = \sigma T^4_{\rm irr}$, we can compute the
irradiation temperature $T_{\rm irr}$ at the surface of the disk as
\begin{equation}
T^4_{\rm irr} =  \frac{L_{\star}}{4 \pi r^2 \sigma} \left( \frac{H}{r} \right) \left(  \frac{d \ln{H}}{d\ln{r}} - 1 \right)   \left( 1 - \beta \right).
\end{equation}

From analytic \citep{Chiang-Goldreich-1997} and numerical solutions
(\cite{D'Alessio-et-al-1998}; \cite{Dullemond-et-al-2002}), it can be shown that the height $H$ of the disk when stellar radiation is
taken into account has a power-law dependence with radius, $H \propto
r^f$, with $f \approx 1.3 - 1.5$. Using these values, we obtain that
$(d\ln{H}/d\ln{r}-1)$ varies slightly between 0.3 to
0.5. The aspect ratio ($H/r$) can reach values from $10^{-4}$ to
$10^{-1}$ (e.g., \cite{Bell-et-al-1997}). Even taking values that give the
maximum temperature from the above equation i.e., $d \ln{H}/d\ln{r} -1
= 0.5$; $H/r = 10^{-1}$; and $\beta =0$ (all radiation penetrates),
and assuming a central star of $ 50 \, \rm L_\odot$ (\cite{Panic-et-al-2010} quotes $26 \, \rm L_\odot$ for HD 100546), the disk
temperature from irradiation at $r = 68 \, \rm au$ reaches $\sim 60 \,
\rm K$, while from viscous dissipation plus radiative feedback the
temperature reaches values $\gtrsim 100 \, \rm K$ depending on the 
planet luminosity $L_{\rm p}$.

For the simulations with a solar-type star presented in \S~\ref{sec:dens}--~\ref{sec:accretionrate}, 
we obtain a surface temperature of about $\sim 55 \, \rm K$ at $r = 10 \, \rm au$
(planet location).  Therefore, the heating of the disk by the central
star is, in the most extreme case, of the same order as the
temperature obtained by viscous dissipation when there is no feedback
($T\sim 55 \rm K$), and one order of magnitude bellow when the
feedback is activated $T\sim 1000 \, \rm K$ (see Figure
\ref{Temp2D-comp}).

We conclude from this that the surface temperature in our 2D disk
model is not expected to be affected by the irradiation energy of the
central star, and that this effect can be neglected from our
calculations.

\subsection{Application to massive black hole binaries}

Even though in this paper we focus on proto-planetary discs, a very
similar physical set-up is that of unequal-mass massive black hole
binaries.  Such binaries are expected to form after galaxy mergers,
once the central massive black hole of each galaxy migrates to the
center of the new system due to dynamical friction (e.g., \cite{Begelman-et-al-1980}).  During the merger, large quantities of gas get funnelled to
the inner region of the new galaxy, where they form a nuclear gas disk
that interacts with the binary (e.g., \cite{Mayer-2007}).  If the
masses of the black holes are dissimilar, the secondary will orbit the
primary, while both are embedded in the gaseous disk, in a situation
very much resembling the star--planet--disk systems we model in this
paper (e.g., \cite{Armitage-Natarajan-2002}), especially considering that
both planets and embedded black holes are expected to produce
luminosity by accretion.

Binaries with separations of parsecs or smaller are not directly
resolvable, so we need to rely on indirect methods to identify them.
One such method is based on the assumption that the gaseous disk
around the primary radiates as a multi-color blackbody.  If the
secondary produces a gap in the disk, then the spectrum will show a
dip at the wavelength range associated with the temperature of the
missing gas (\cite{Chang-et-al-2010}; \cite{Gultekin-Miller-2012}).
Extrapolating our results to that regime, it is clear that the disk
spectra will also be modified by the heating up of the gas surrounding
the secondary black hole, likely producing a larger dip and a
concurrent increase of the shorter wavelength emission.  A more
detailed analysis of this situation is deferred to a follow-up paper.

\section{Summary}\label{summary}

In this paper we present 2D hydrodynamical simulations of the
interaction between a Jupiter-mass planet and its parent
protoplanetary disk, including for the first time the effect of the
planet's radiative feedback onto the disk (energy released by the
planet as it assembles its mass). In this first work, we assume that
the energy released by the planet to the disk is constant over the
duration of our simulations, a few hundred planet orbits typically.
We carried out various simulations taking realistic parameters for
protoplanetary disks that could be identified with actually observed
systems, like e.g., HD 100546 (\cite{Currie-et-al-2014}, \cite{Quanz-et-al-2013b}), varying the luminosity of the planet.

We find that planet luminosities below $L_{\rm p} \lesssim 10^{-4} \rm
L_\odot$ barely modify the disk response to the planet: the gap formed
by the planet does not show any noticeable modification compared to
without feedback and the emitted spectrum of the disk remains
practically unchanged.  However, planets with $L_{\rm p} \gtrsim
10^{-4} \rm L_\odot$ introduce significant changes: the additional
energy input from the planet heats up the disk, locally enhancing the
local disk accretion rate (understood into the primary), increasing
even more the temperature at this location. This mechanism results in
a positive feedback magnifying the energy output until a thermal
balance is reached.

Figure \ref{Acc_gap} shows the increase in the disk accretion rate in the
planet vicinity. The spectrum and luminosity of the disk are modified,
shifting the emissions to higher amplitudes and shorter wavelengths.
For instance, assuming $10^{-3} \, \rm L_\odot$ for the planet
feedback, the accretion rate at the planet vicinity increases from $6
\times 10^{-8}$ (without feedback) to $10^{-6}$ $\rm M_\odot\,
yr^{-1}$, and translates into an increase in the disk luminosity from
$ L_{\rm disk} = 5.5 \times10^{-2} \, \rm L_\odot$ (without feedback;
$\lambda_{\rm peak} = 19.3 \mu m$) to $L_{\rm disk} = 0.9 \, \rm
L_\odot$ ($\lambda_{\rm peak} = 2.5 \mu m$).

We find that our results do not depend on the viscosity
prescription. Thus, in our models, setting the planet luminosity will
fix the accretion rate, temperature and luminosity in a region close
to the planet.  Our results imply that observations of protoplanetary
disks where planets are form could reveal the accretion process onto
them, without much interference from nuance parameters such as the
viscosity.

Finally, we build a model for the system around HD 100546 \citep{Quanz-et-al-2013b}, reproducing quite well the observed flux density in
$L^\prime$ band of 1.3 mJy from the region around the accreting
candidate planet HD 100546b. Our model indicates that this system
contains a forming planet with a luminosity of $\sim 10^{-4} \, \rm
L_\odot$.

\section*{Acknowledgments}

We thank Alberto Sesana for illuminating discussions when starting
this project, Phil Armitage for very useful comments on an earlier
version of this paper, and Christoph Mordasini for fruitful discussions. 
We also thank students Gabriel Torrealba and
Mat\'ias G\'arate for their help in developing analysis and
visualization tools for Fargo simulations.  MM acknowledges support
from FONDECYT grant No. 3120101 and CONICYT-Gemini grant No. 32130007, JC acknowledges support from
CONICYT-Chile through FONDECYT (1141175), Basal (PFB0609), Anillo
(ACT1101) and DRI-Intercambio (PCCI130064) grant, SP acknowledges
financial support provided by FONDECYT grant 3140601, SC acknowledges
support from Millennium Science Initiative, Chilean Ministry of
Economy: Nucleus P10-022-F, and from FONDECYT grant 1130949.
M.M., J.C., S.P., and S.C. acknowledge financial support from 
Millennium Nucleus RC130007 (Chilean Ministry of Economy).

\bibliography{astro}

\end{document}